\documentclass[apj]{emulateapj}
\usepackage{color}
\usepackage{rotating}
\usepackage{multirow}

\begin{document}

\title{A Comparative Study of Long and Short GRBs. II. A Multi-wavelength Method to distinguish Type II (massive star) and Type I (compact star) GRBs}

\author{Ye Li$^{1,2,3}$, Bing Zhang$^{3}$, Qiang Yuan$^{4,5,6}$}

\affil{
$^1$Kavli Institute for Astronomy and Astrophysics, Peking University, Beijing 100871, China; liye$\_$kiaa@pku.edu.cn \\
$^2$National Astronomical Observatories, Chinese Academy of Sciences, Beijing 100012, China \\
$^3$Department of Physics and Astronomy, University of Nevada, Las Vegas, NV 89154, USA; zhang@physics.unlv.edu \\
$^4$Key laboratory of Dark Matter and Space Astronomy, Purple Mountain Observatory, Chinese Academy of Sciences, Nanjing 210008, Peopleʼs Republic of China; yuanq@pmo.ac.cn \\
$^5$School of Astronomy and Space Science, University of Science and Technology of China, Hefei 230026, Anhui, Peopleʼs Republic of China \\
$^6$Center for High Energy Physics, Peking University, Beijing 100871, China
}

\begin{abstract}

Gamma Ray Burst (GRBs) are empirically classified as long-duration GRBs (LGRBs, $>$ 2s) and short-duration GRBs (SGRBs, $<$ 2s). Physically they can be grouped into two distinct progenitor categories: those originating from collapse of massive stars (also known as Type II) and those related to mergers of compact stars (also known as Type I). Even though most LGRBs are Type II and most SGRBs are Type I, the duration criterion is not always reliable to determine the physical category of a certain GRB. Based on our previous comprehensive study of the multi-wavelength properties of long and short GRBs, here we utilize the Naive Bayes method to physically classify GRBs as Type I and Type II GRBs based on multi-wavelength criteria. It results in 0.5\% training error rate and 1\% test error rate. Moreover, there is a gap [-1.2, -0.16] in the distribution of the posterior Odds, $\log O({\rm II:I})$, the Type II to Type I probability ratio. Therefore, we propose to use ${\cal O} = \log O({\rm II:I})+0.7$ as the parameter to classify GRBs into Type I ($<0$) or Type II ($>0$). The only confirmed Type I GRB, GRB 170817A, has log $O({\rm II:I})=-10$. According to this criterion, the supernova-less long GRBs 060614 and 060505 belong to Type I, and two controversial short GRBs 090426 and 060121 belong to Type II.

\end{abstract}

%
%
%

\section{Introduction}
Gamma Ray Bursts (GRBs) are intense 
bursting $\gamma$-ray emitting events in the universe. Multi-wavelength, multi-messenger observations over the years suggest that they can be broadly classified into two physically distinct categories, those originating from core collapse of massive stars  
\citep{1993ApJ...405..273W, 1998ApJ...494L..45P, 1999ApJ...524..262M},
and those originating from mergers of compact stars, e.g., neutron star - neutron star (NS-NS)
or neutron star - black hole (NS-BH) mergers (\cite{1986ApJ...308L..43P,
1989Natur.340..126E,1992ApJ...395L..83N}, see \cite{2014ARA&A..52...43B} for a review).
These two physically distinct types are also termed as Type II (massive star) GRBs and Type I (compact star) GRBs \citep{2006Natur.444.1010Z,2009ApJ...703.1696Z}.

This physical classification scheme of GRBs { is} generally consistent with the phenomenological classification of GRBs based on the prompt $\gamma$-ray durations
\citep{1993ApJ...413L.101K}. Type II GRBs typically have $\gamma$-ray durations ($T_{90}$) longer than 2 s (LGRBs), while Type I GRBs typically have $T_{90}$ shorter than 2 s\footnote{ In the literature, an ``intermediate'' duration type between short and long GRBs has been studied phenomenologically  as a statistically significant population based mainly on the $T_{90}$ criterion \citep{1998ApJ...508..314M, 1998ApJ...508..757H, 2003ApJ...582..320H,
2006A&A...447...23H, 2018Ap&SS.363...53H, 2013PASJ...65....3T}. No evidence suggests that they form a physically distinct category. They may be a consequence of an instrumental effect \citep{2003ApJ...582..320H} or may belong to a subclass of Type II GRBs GRB \citep{2013ApJS..209...20G}.}
This connection is rooted from the ``density argument'' \citep[e.g.][]{Zhang2018book}, i.e., the average density of massive star is much smaller than that of compact stars, so that the free-fall and the accretion time scale of the former is much longer than that of the { latter}. Observationally, the following multi-wavelength observational properties all point towards a connection between long and Type II GRBs and between short and Type I GRBs.  LGRB host galaxies are usually dwarf galaxies with a high star formation rate and low metallicity \citep{1997Natur.387..476S, 1998ApJ...507L..25B, 2002AJ....123.1111B, 2002ApJ...566..229C, 2004A&A...425..913C, 2009ApJ...691..182S, 2015A&A...581A.125K}. Typically LGRBs are located in bright regions of their hosts, with small offsets from the galaxy center \citep{2002AJ....123.1111B, 2006Natur.441..463F, 2016ApJ...817..144B}. The smoking-gun signature connecting LGRBs with massive star core collapse is the direct detection of a Type Ic supernova (SN) \citep{1998Natur.395..670G, 2003Natur.423..847H, 2003ApJ...591L..17S, 2006ARA&A..44..507W, 2012grbu.book..169H, 2013ApJ...776...98X} associated with a LGRB at least in some nearby events, and data are consistent with the majority of LGRBs have such an association (even though in most cases, the SN is not easy { to detect}).
The connection between short and Type I GRBs is supported by the diversity of the short GRB host galaxies, from dwarf to elliptical galaxies \citep{2005Natur.437..851G, 2005Natur.438..988B}. Within the hosts, SGRBs are usually located in the faint regions with large offsets from the center \citep{2010ApJ...708....9F, 2011ApJ...734...96K, 2013ApJ...776...18F} with a small local specific star formation rate. This is consistent with the expected delay between star formation and the merger of the two compact stars that give rise to the GRB. No SN was found to be associated with SGRBs, with very stringent non-detection limits \citep{2005Natur.437..845F, 2005ApJ...630L.117H, 2005Natur.437..859H, 2011ApJ...734...96K, 2013ApJ...774L..23B,2014ARA&A..52...43B}. A few SGRBs are found to be associated with ``kilonova/macronova' events 
\citep{1998ApJ...507L..59L, 2010MNRAS.406.2650M, 2013Natur.500..547T, 2013ApJ...774L..23B, 2015NatCo...6E7323Y, 2015ApJ...807..163G, 2016arXiv160307869J}, which lends support to the neutron star merger scenario. Finally, the direct discovery of the gravitational wave event GW170817 and its associated SGRB 170817A \citep{2017PhRvL.119p1101A, 2017ApJ...848L..13A, 2017ApJ...848L..14G,2018NatCo...9..447Z} and kilonova AT1027gfo \citep{2017Sci...358.1556C, 2017ApJ...848L..12A,2017ApJ...851L..21V} firmly established the Type I - origin of at least some SGRBs.

While $T_{90}$ is widely used to define the physical origin of a GRB, it is not always reliable. For example, GRB 060614 was a famous long GRB without a supernvoa explosion. Its prompt $\gamma$-ray emission has a 4.5 s spike with extended emission lasting for longer than 100 s. However, its spectral lag is very short \citep{2006Natur.444.1044G, 2010ApJ...717..411N}. When scaling down in energy, its observational properties are similar to those of a SGRB with extended emission \citep{2007ApJ...655L..25Z}. It is located in a faint region of a passive host galaxy \citep{2006Natur.444.1053G, 2006Natur.444.1047F, 2016ApJ...817..144B}. A very stringent upper limit was placed against its association with any SN \citep{2006Natur.444.1053G, 2006Natur.444.1050D, 2006Natur.444.1047F}. Later, a putative kilonova was reported \citep{2015NatCo...6E7323Y}. All these suggest that it very likely has a Type I origin. Another example is the short GRB 090426, which has a $T_{90}$ of 1.24 s. Phenomenologically, it belongs to the SGRB category. However, it is located in the central region of a blue interacting host galaxy, which is more consistent with a Type II-origin \citep{antonelli2009, 2010MNRAS.401..963L}. Its amplitude parameter $f$ { (the ratio between the peak flux and the background flux)} is small, suggesting that there is a high probability that the observed short duration is simply the tip-of-iceberg of a long-duration GRB \citep{2014MNRAS.442.1922L}.  
The only confirmed Type I GRB 170817A belongs to the fainter and softer { category} in the phenomenological short GRB sample \citep{2018NatCo...9..447Z, 2018MNRAS.481.1597G}.
Without GW association, this burst was { unremarkable} among faint short GRBs detected by Fermi/GBM. It became unique when the distance information (and hence, its extremely low luminosity and energy) was revealed. Compared with traditional short GRBs, its special properties are likely related to the viewing angle effect.
In general, the $T_{90}$ information may be misleading at least for some bursts, and multi-wavelength data are essential to diagnose the physical category of  GRBs.

It has been suggested that one should apply multiple observational criteria, including both prompt emission and host galaxy information, to define the physical category of a GRB \citep{2009ApJ...703.1696Z}.
{ However, no quantitative method has been proposed to carry out this task. 
Thanks to the larger sample and more available information, we now enter the era of Astroinformatics and Astrostatistics, and are able to apply elaborate multivariate classification methods \citep{1998ApJ...508..314M, 2019BAAS...51c.355S, 2011ApJS..194....4B, 2013ApJS..209...32B}\footnote{https://asaip.psu.edu/}. 
To perform multivariate classifications, a large field of mathematical methods, such as Logistic Regression, Naive Bayes, Support Vector Machines, and Artificial neural network, { and other methods} have been developed \citep{James2013, machinelearning}\footnote{https://see.stanford.edu/Course/CS229}. 
In this paper, we choose to use the Naive Bayes method due to its simplicity, understandability, and the limited size of the GRB sample.
}
The method is presented in Section 2 and the results are given in Section 3.
Section 4 presents the conclusions with some discussion.

\section{Naive Bayes Method}
Naive Bayes classifiers are a group of Baysian-theorem-based classification methods\footnote{https://en.wikipedia.org/wiki/Naive\_Bayes\_classifier.}.
It is simple, fast, and one of the most understandable classifier in machine learning.
According to the Bayesian theorem, the posterior probability of one GRB
with parameters $\{x\}=\{x_1,x_2,...,x_i\}$ to be a hypothesized Type is
\begin{equation}
P({\rm Type}|\{x\}) = \frac{P(\{x\}|{\rm Type})P({\rm Type})}{P(\{x\})},
\label{nb}
\end{equation}
where $P({\rm Type})$ is the prior probability of a GRB to be one specific type,
$P(\{x\})$ is the probability of the parameter set $\{x\}$, and $P(\{x\}|{\rm Type})$
is the likelihood of one specific type of GRB to have a parameter set $\{x\}$.
{ ``Naive'' Bayes assumes parameters are independent, thus, $$P(\{x\}|{\rm Type})=\prod\limits_{i}P(x_i|{\rm Type}).$$
Although this assumption is strong, it turns out that Naive Bayes performs surprisingly well even if the parameters are mildly correlated
\citep{hand2001, 2011ApJS..194....4B}. 
The implementation of Naive Bayes usually follows the following steps: \begin{enumerate}{\itemindent=3em}
    \item Estimate the likelihood $P(x_{\rm i}|{\rm Type})$ for each parameter $x_{\rm i}$ with the preliminary Type I and Type II GRB samples; 
    \item Missing values imputation -- replace the missing values by substituted values, see Section 2.2 for more information;
    \item Estimate the priors $P({\rm I})$ and $P({\rm II})$, usually by the size of each sample; 
    \item Calculate the posterior probabilities following equation (\ref{nb});
    \item Normalize the posterior probabilities $P({\rm I}|\{x\})$ and $P({\rm II}|\{x\})$ by requiring $P({\rm I}|\{x\})+P({\rm II}|\{x\})=1$ for each object;
    \item The type of GRB is assigned to the one with a higher posterior probability.
\end{enumerate}
The likelihood for each parameter $P(x_i|{\rm Type})$ is estimated with the observed sample in Section \ref{chp_pdistribution}. The parameter selection and the priors are presented in Section \ref{chp_classification}.}

\subsection{Statistical distributions of GRB observational properties \label{chp_pdistribution}}
Our method makes use of the multi-wavelength data of GRBs. In order to come up with a set of sound classification criteria, one needs to first look into the statistical distributions of the observational properties for different types of GRBs. We use a sample based on the catalog presented in \cite{2016ApJS..227....7L}, in which the prompt emission and host galaxy parameters of 407 GRBs were compiled.
In total, there are 16 parameters, including the redshift, 7 prompt emission parameters, and 8 host galaxy parameters. The prompt emission parameters include the duration $T_{90}$, two spectral parameters (the peak energy $E_{\rm p}$ and the low energy photon index $\alpha$) of the best-fitting Band function \citep{1993ApJ...413..281B}, the isotropic $\gamma$-ray energy $E_{\rm iso}$, the isotropic $\gamma$-ray peak luminosity $L_{\rm iso}$, the amplitude $f$ parameter, and the effective amplitude parameter $f_{\rm eff}$ (the $f$ parameter when the background is shifted to make the duration $T_{90}$ be 2 seconds, see \citealt{2014MNRAS.442.1922L} for detailed discussion of $f$ and $f_{\rm eff}$ parameters). The host galaxy parameters include stellar mass of the host $M_*$, star formation rate (SFR), specific star formation rate (sSFR; the ratio between SFR and $M_*$), metallicity of the host [X/H], half-light radius of the host $R_{50}$, offset of the GRB from the center of the host galaxy in units of kpc $R_{\rm off}$, normalized offset $r_{50}=R_{\rm off}/R_{50}$, and fraction of the host light in the area fainter than the GRB position, $F_{\rm light}$.

The ``consensus'' LGRBs and SGRBs are defined based on the online GRB catalog maintained by Jochen Greiner\footnote{http://www.mpe.mpg.de/$\sim$jcg/grbgen.html}. 
{ The LGRBs and SGRBs in this catalog are defined based on the published results in the literature, including GCNs.}
Such definitions { usually} take into account the multi-wavelength observational data, and the definitions of ``long'' and ``short'' already implies their possible physical origins. { There are no well-defined criteria regarding how the category of one particular burst is assigned, but the classification presented on the website usually reflects the consensus in the GRB community regarding each GRB with redshift and multi-wavelength information. }
We consider these ``consensus'' LGRB sample as the preliminary sample of Type II GRBs, and the ``consensus'' SGRB sample as the preliminary sample of Type I GRBs. { The classification of some bursts are subject to debate, e.g. GRB 060505, GRB 060614, GRB090426, and  GRB 060121. These bursts are excluded from our control sample. Altogether,  there are 403 GRBs in our control sample.}

For each parameter { except $F_{\rm light}$}, we use a Gaussian function
\begin{equation}
P(x_i|{\rm Type})=\frac{1}{\sqrt{2\pi \sigma^2}}{\rm exp}\big( - \frac{(x_i-\mu)^2}{2\sigma^2}\big)
\end{equation}
to fit the distribution of each of these two classes of GRBs. 
{ For $F_{\rm light}$, an exponential distribution 
\begin{equation}
P(x_i|{\rm Type})=\frac{\gamma}{{\rm exp}(\gamma)-1}e^{\gamma x_i}, 
\end{equation}
is used to fit the distribution.
The normalization is obtained by requiring the integration in $[0,1]$ to be unity.}

For most of the parameters, not all GRBs have the measured values. We properly select a smaller sample to perform the fittings. For the redshift parameter, we only consider the precise, spectrally identified redshifts. The original definition of $f_{\rm eff}$ \citep{2014MNRAS.442.1922L} is slightly larger than 1. In this work, we use log ($f_{\rm eff}-1$) instead to perform the statistical analysis. The stellar mass of the host galaxy $M_*$ can be estimated by either the spectral energy density (SED) fitting or the infrared (IR) luminosity. The IR luminosity method assumes a 70-Myr old stellar population, which gives a very large uncertainty. Therefore, we only select the $M_*$ values that are derived using the SED method. The star formation rate (SFR) can be estimated by emission lines such as H$\alpha$, H$\beta$,[OIII], [OII], Ly$\alpha$, as well as continuum such as UV, IR, and SED fitting. Emission lines indicate the SFR with age 0-10 Myr, around the life of stars with mass $> 30\ \rm M_{\odot}$. The continuum, on the other hand, indicates the SFR with age 0-100 Myr \citep{2012ARA&A..50..531K}. In our analysis, we only adopt the values obtained from the emission lines to perform the fitting. We further exclude the upper and lower limits of the SFR. The metallicity [X/H] estimated with $R_{23}=\rm ([OII]\lambda3727+[OIII]\lambda4959, 5007)/H\beta$ is double-valued \citep{2008ApJ...681.1183K, 2009ApJ...691..182S}. Following the suggestions of \cite{2004ApJ...617..240K} and \cite{2009ApJ...690..231B}, we use the larger one of the two values as the metallicity of a particular GRB.

\begin{figure*}[!b]
\centering
\includegraphics[width=0.35\textwidth]{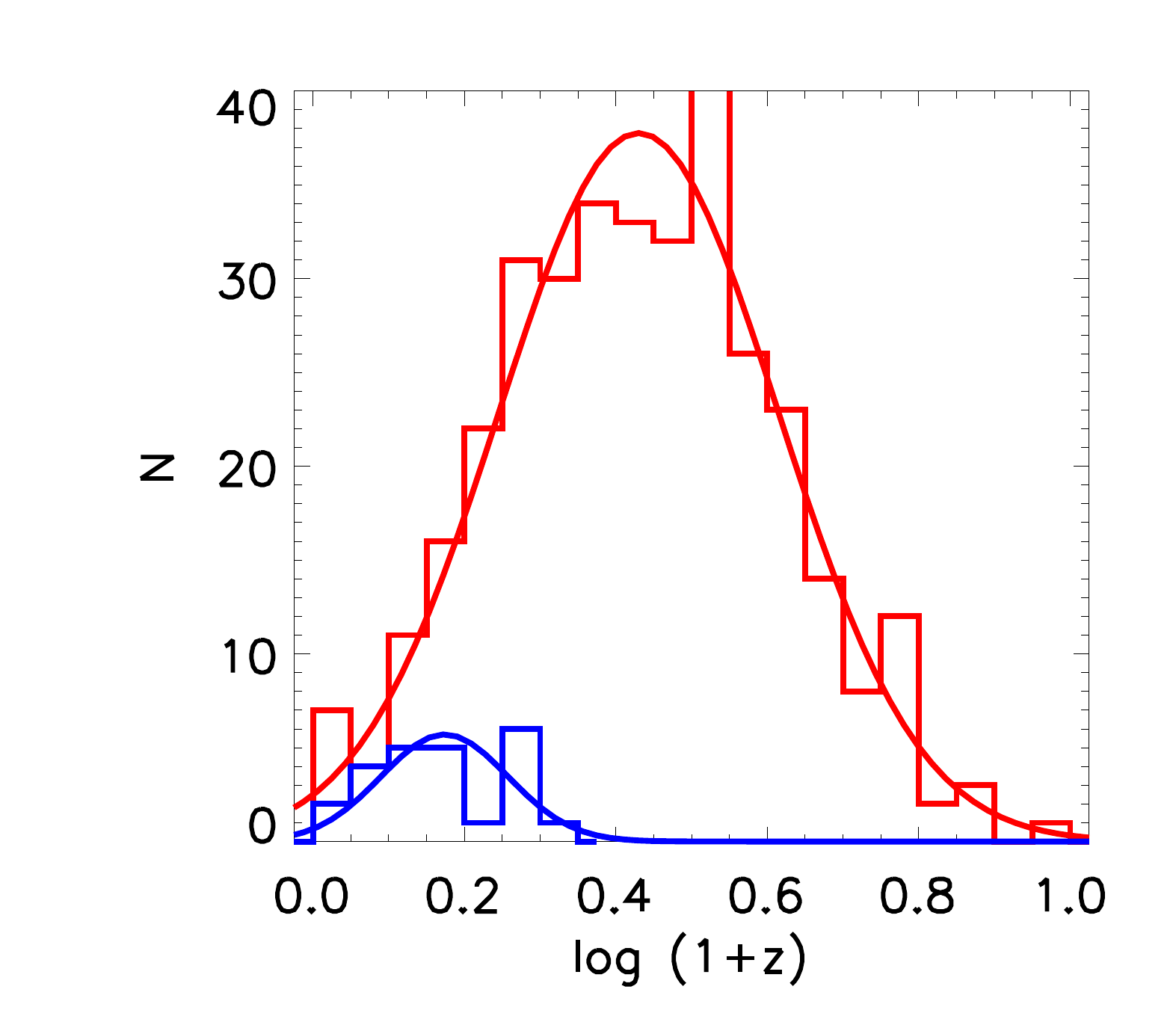}
\includegraphics[width=0.35\textwidth]{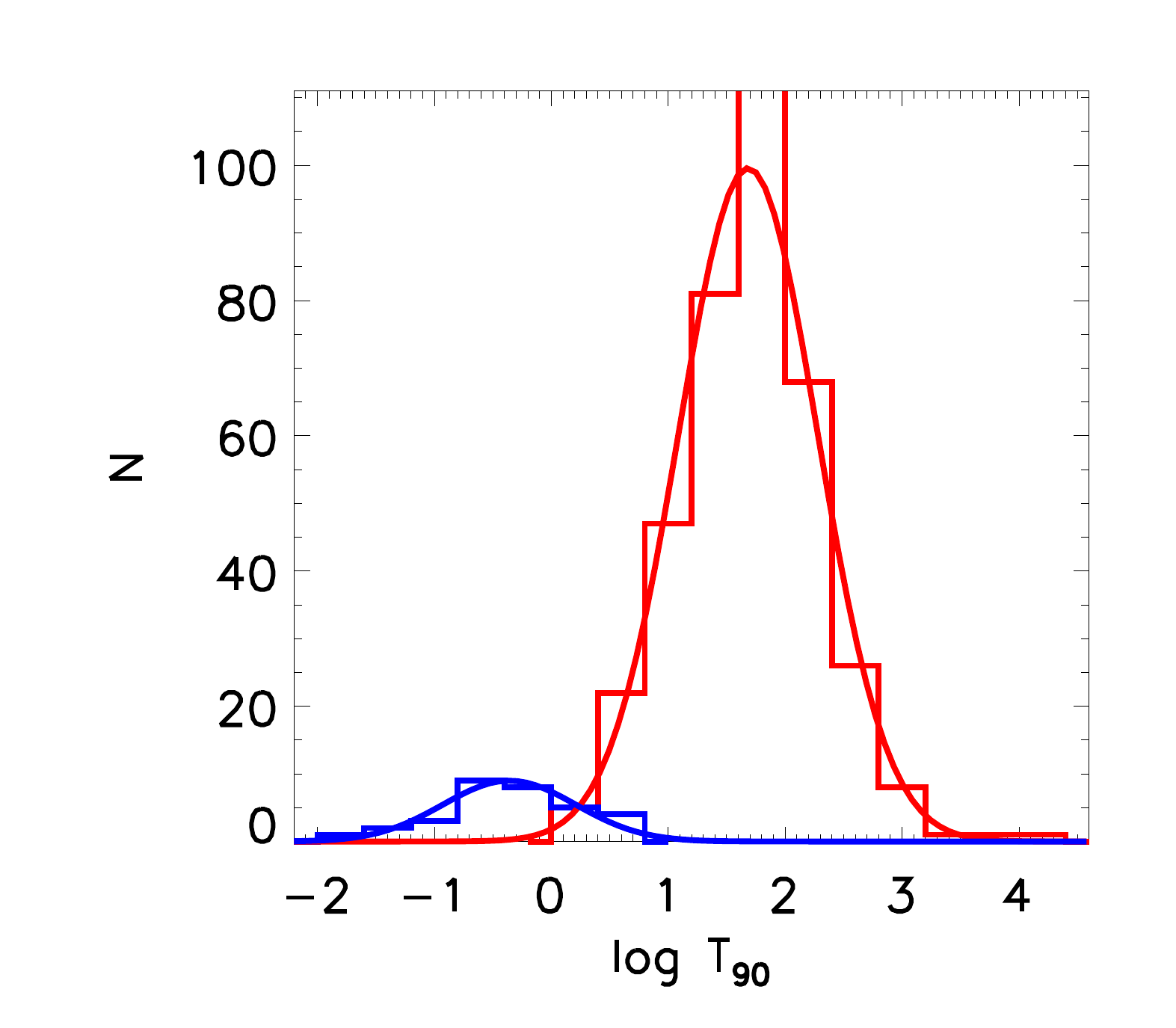}
\includegraphics[width=0.35\textwidth]{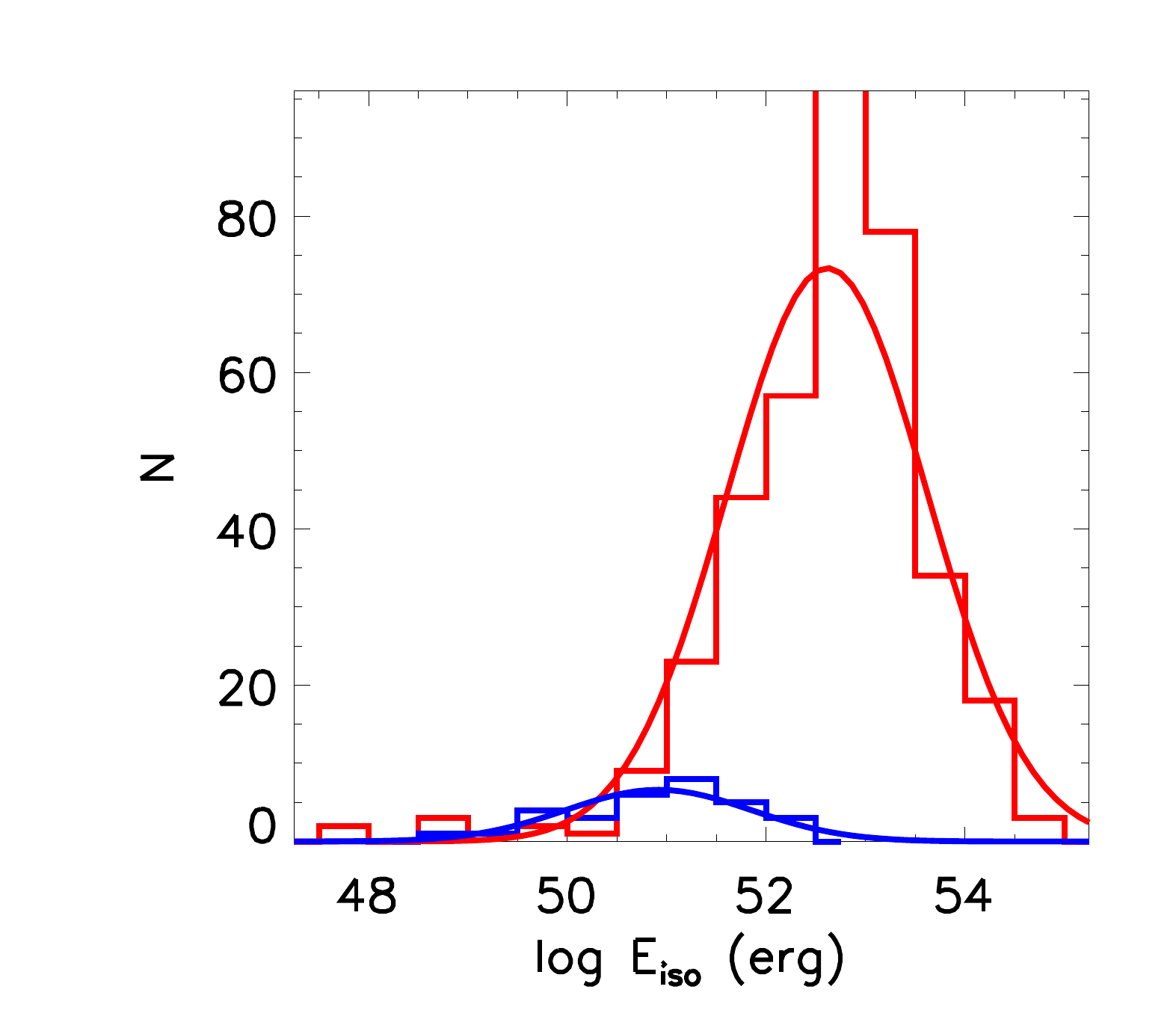}
\includegraphics[width=0.35\textwidth]{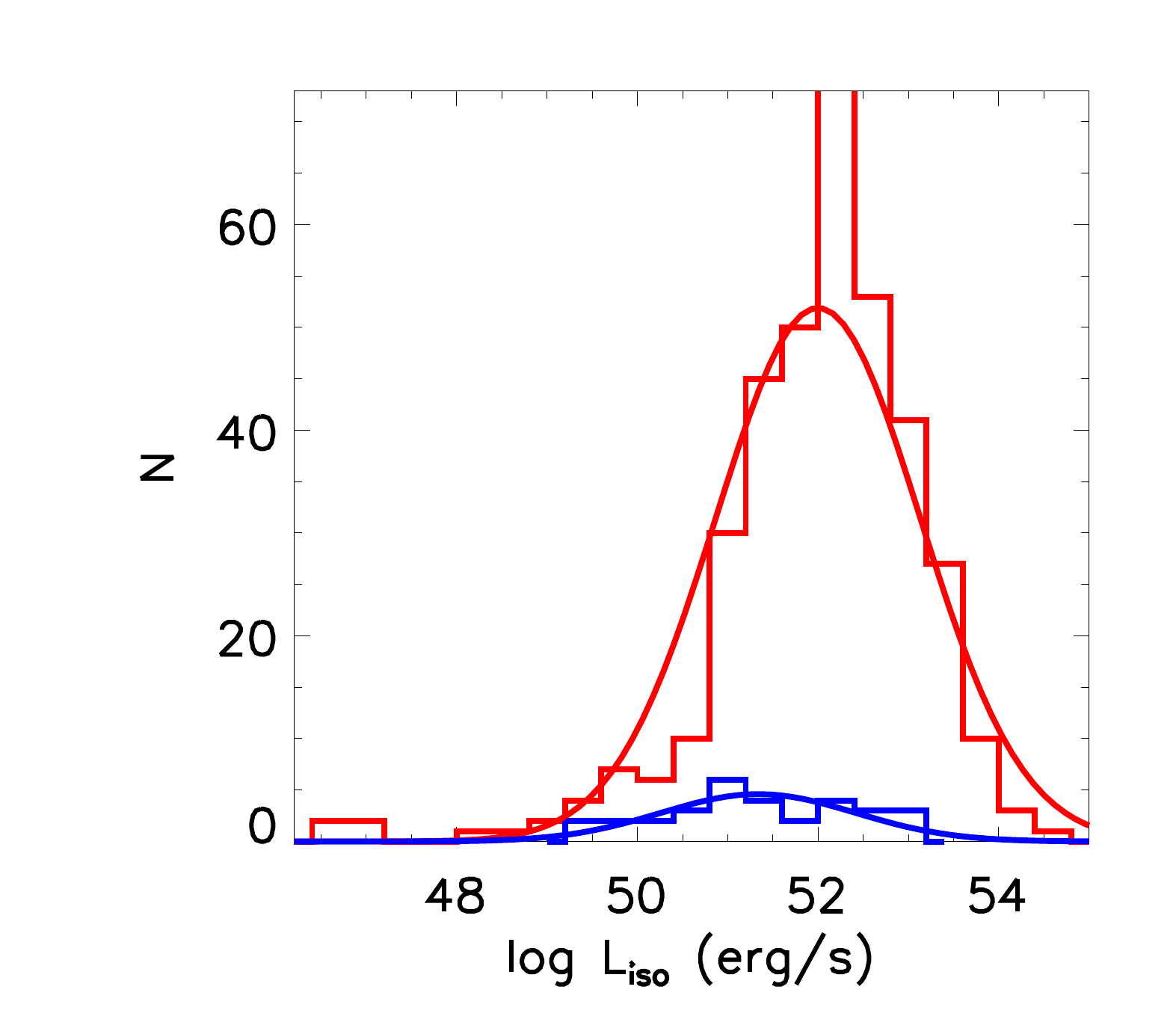}
\includegraphics[width=0.35\textwidth]{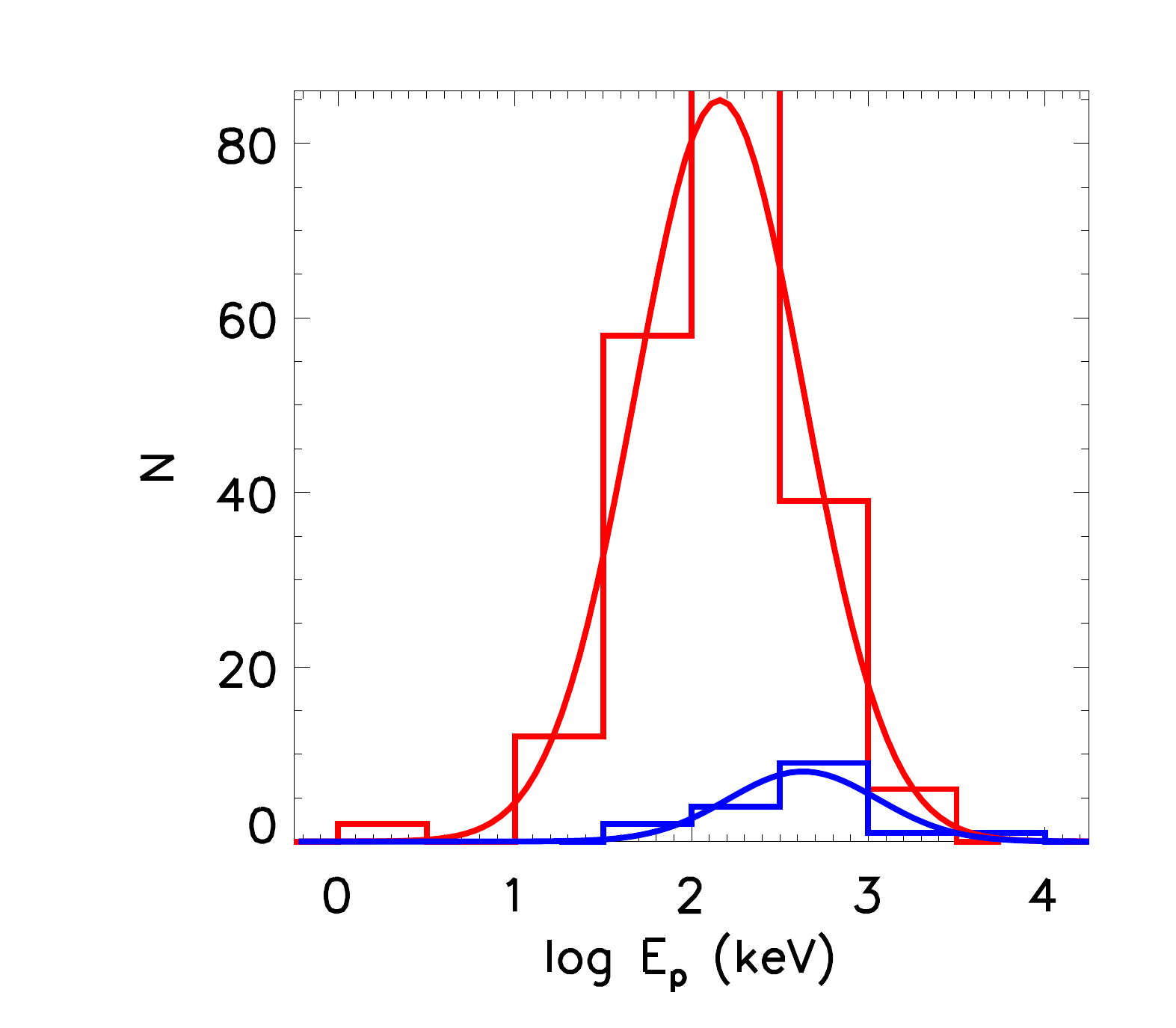}
\includegraphics[width=0.35\textwidth]{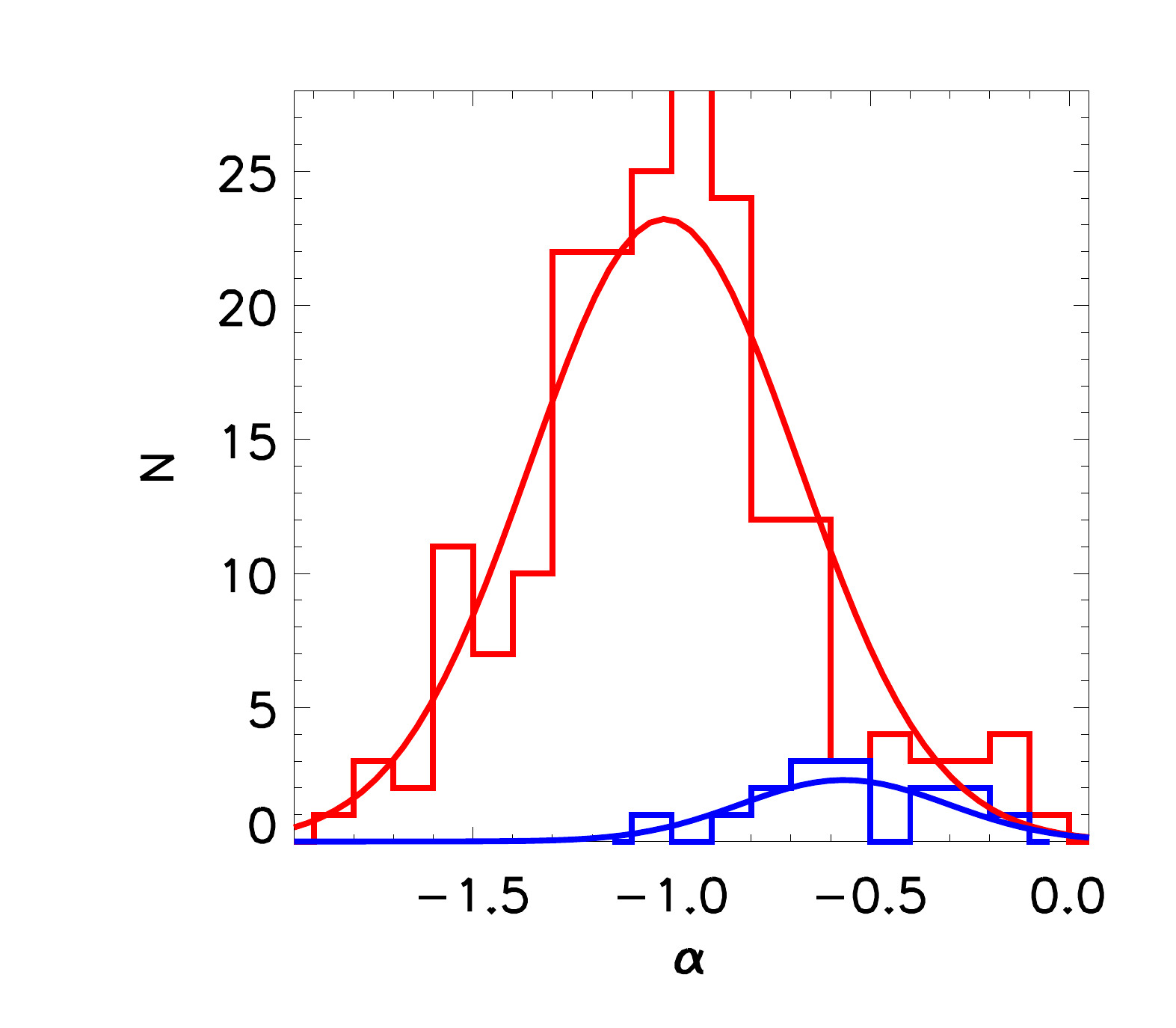}
\includegraphics[width=0.35\textwidth]{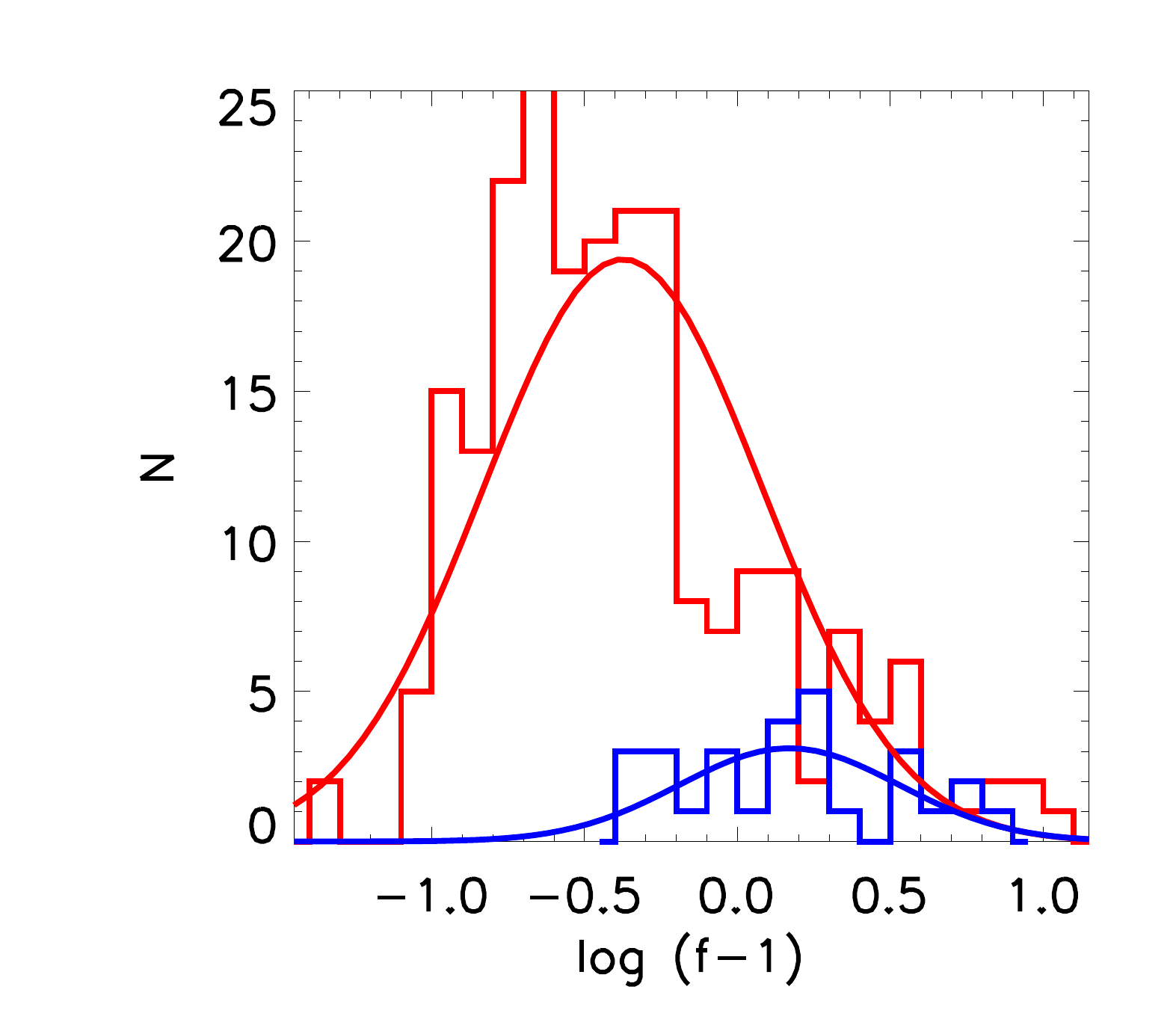}
\includegraphics[width=0.35\textwidth]{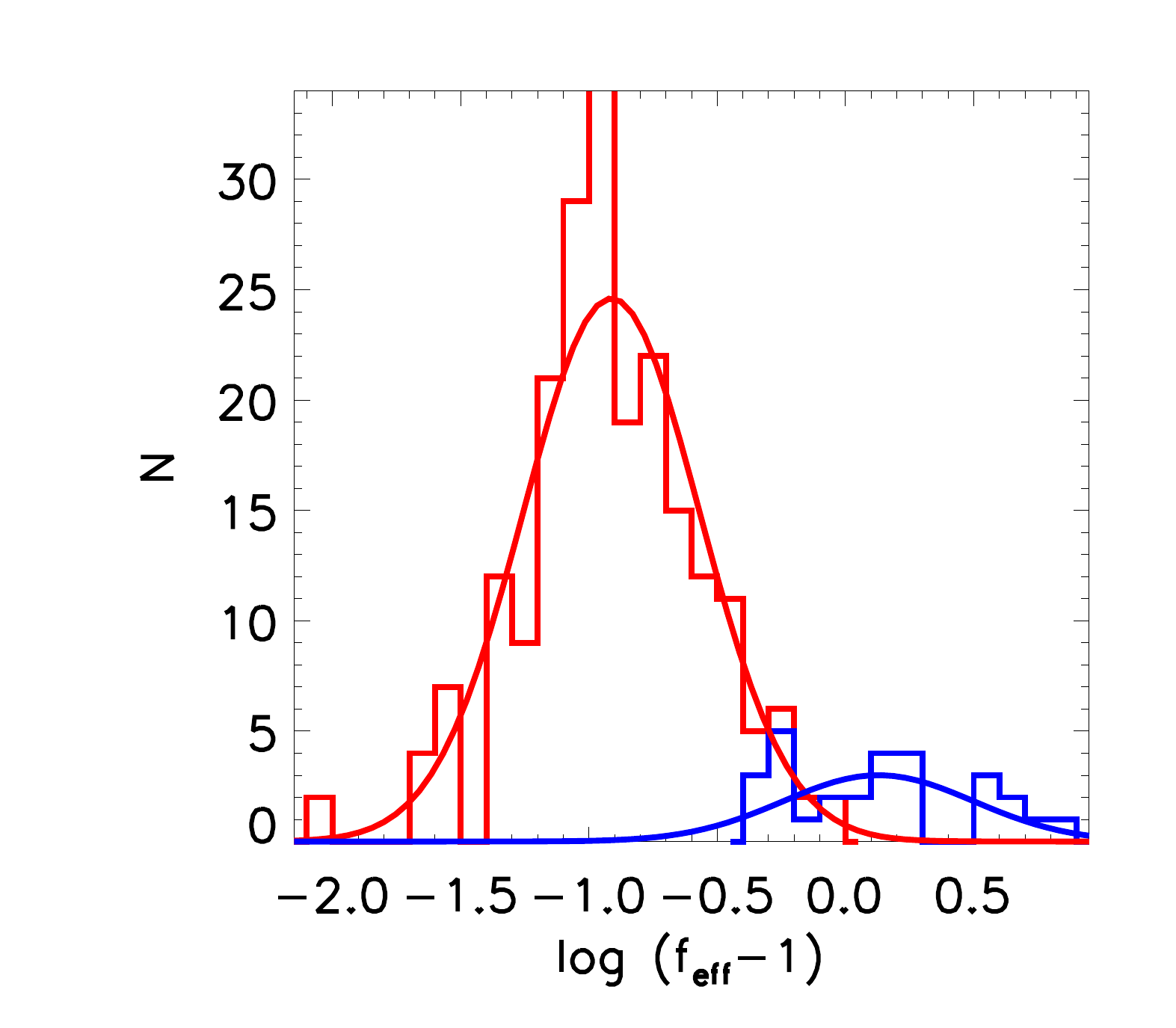}

\caption{
Distributions of prompt and host galaxy parameters of preliminary Type II GRBs (LGRBs; red histograms)
and Type I GRBs (SGRBs; blue histograms), with fitted Gaussian functions (red and blue solid lines) overplotted.
}
\label{figfit}
\end{figure*}

The maximum likelihood estimation is used to fit the distributions. 
For Gaussian distribution, there are analytical solutions of $\mu$ and $\sigma^2$. (See Appendix for the details.)
The histogram of each parameter
and the best fitting results are shown in Fig.~\ref{figfit},
with red lines for Type II GRBs and blue lines for Type I GRBs.
The fitting results are given in Table \ref{tbfit}, with errors estimated by 100 bootstraps. 
The number of samples $N$, mean $\mu$ and standard deviation $\sigma$ for Type II and Type I GRBs are listed in Columns $2-4$ and $7-9$ of Table \ref{tbfit}, for all parameters except $F_{\rm light}$. 
For $F_{\rm light}$, Column 3 and 8 give the exponential index $\gamma$. 
{ In column 12, the values with the same probability in Type I and Type II are presented as the ``imputer'', which are the values to fill in if the corresponding values are missing. }
{ In order to test the goodness of the fit, we applied the Kolmogorov-Smirnov (KS) test, because it is valid for most functions and more sensitive to the center{ (see more discussions about Anderson-Darling test in Section 4.4)}. The KS test results $D_{\rm KS}$ and the corresponding null probability $P_{\rm KS}$ are listed in columns 4 and 5, 8 and 9. For the $F_{\rm light}$ of Type I GRB, the large $D_{\rm KS}$=0.56 is due to many Type I GRBs with $F_{\rm light}=0.0$. However, it is highly influenced by the rounding in the literature. We thus try to randomly assign $F_{\rm light}=0.0$ to be [0.0, 0.05]. It results in $D_{\rm KS}=0.34$ and $P_{\rm KS}=0.02$. If we assign $F_{\rm light}=0.0$ to be [0.0, 0.1]}, $D_{\rm KS}=0.25$ and $P_{\rm KS}=0.19$. The classification results are not affected by the rounding effect correction.

\begin{figure*}
\centering
\includegraphics[width=0.36\textwidth]{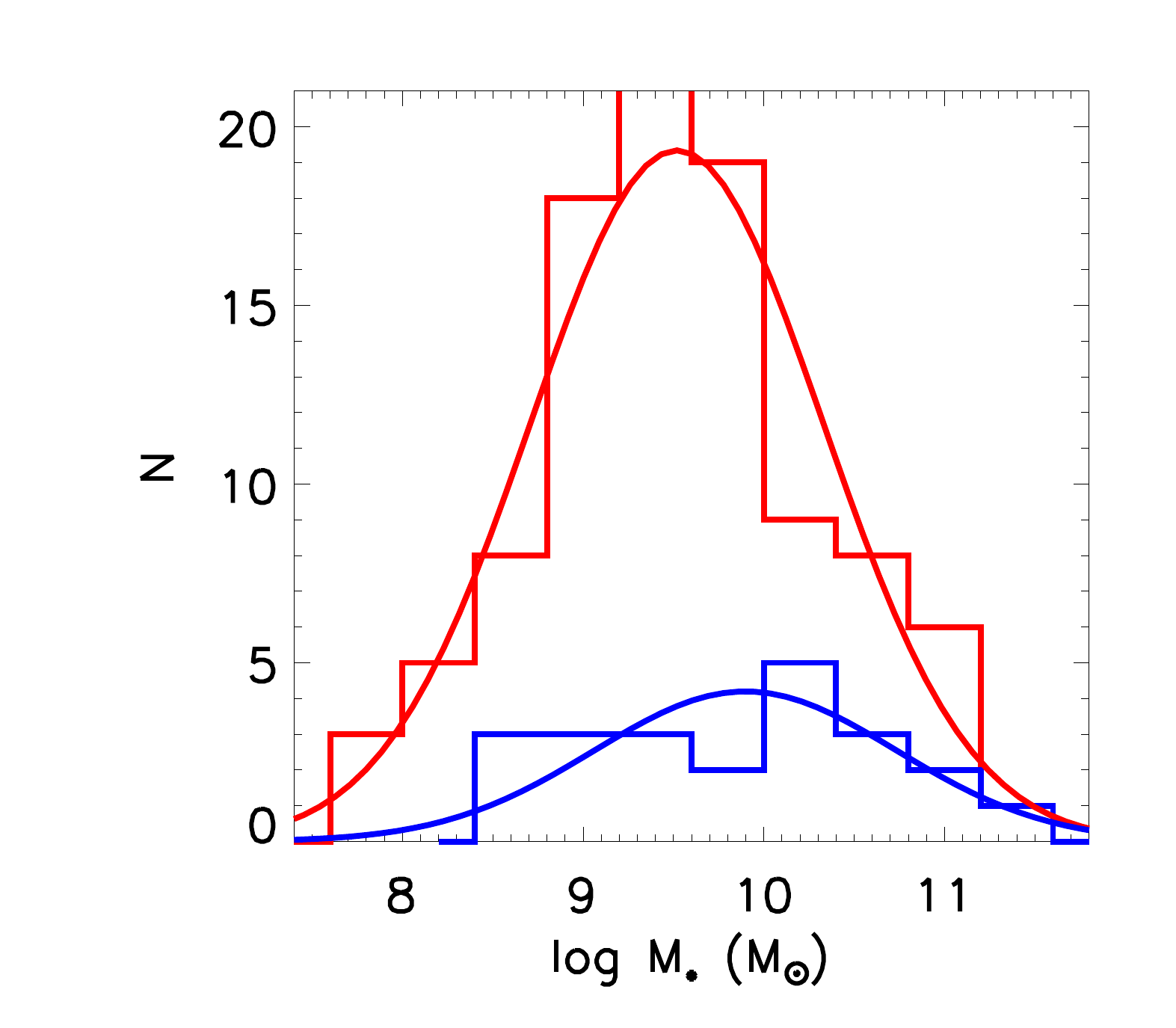}
\includegraphics[width=0.36\textwidth]{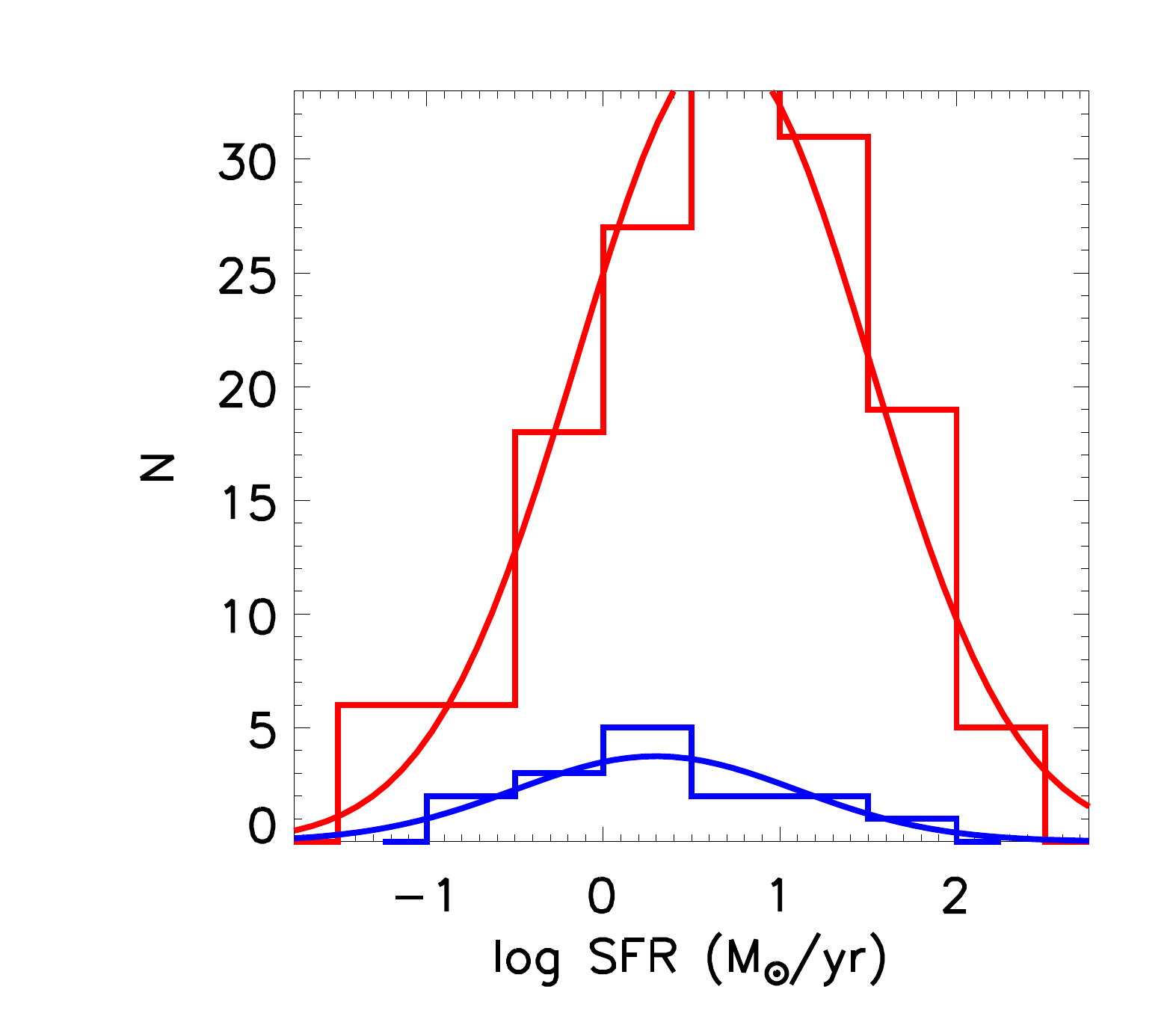}
\includegraphics[width=0.36\textwidth]{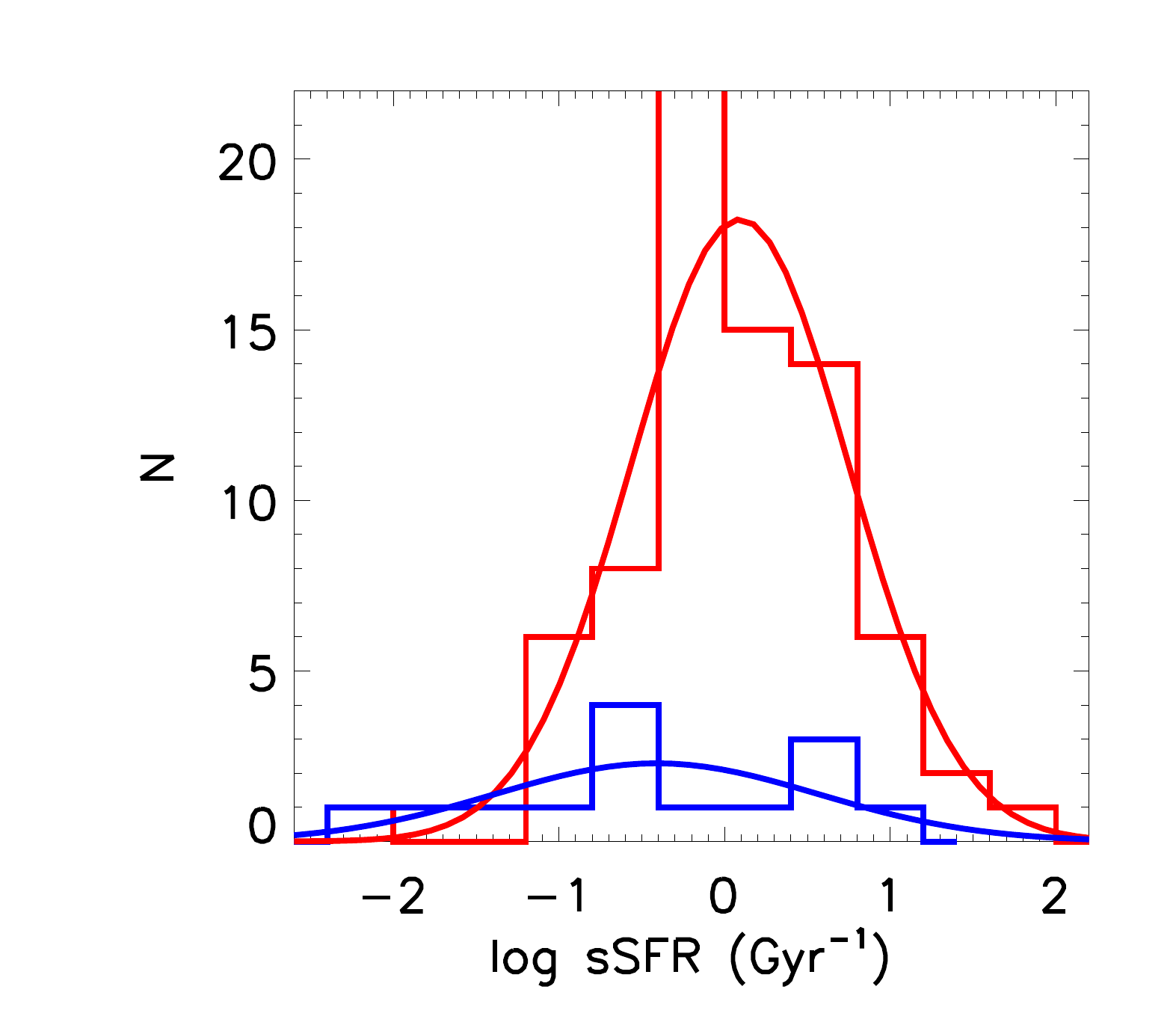}
\includegraphics[width=0.36\textwidth]{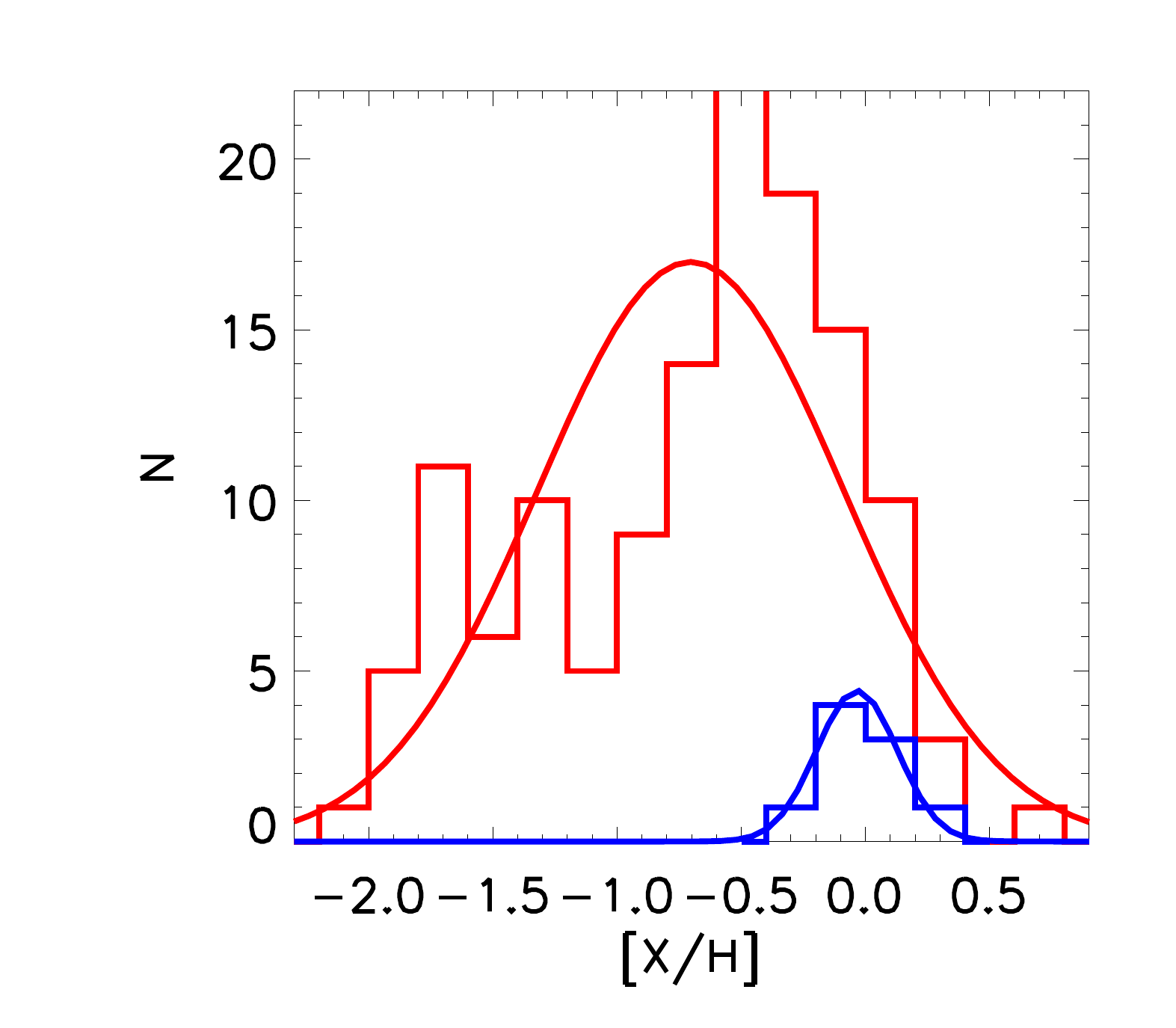}
\includegraphics[width=0.36\textwidth]{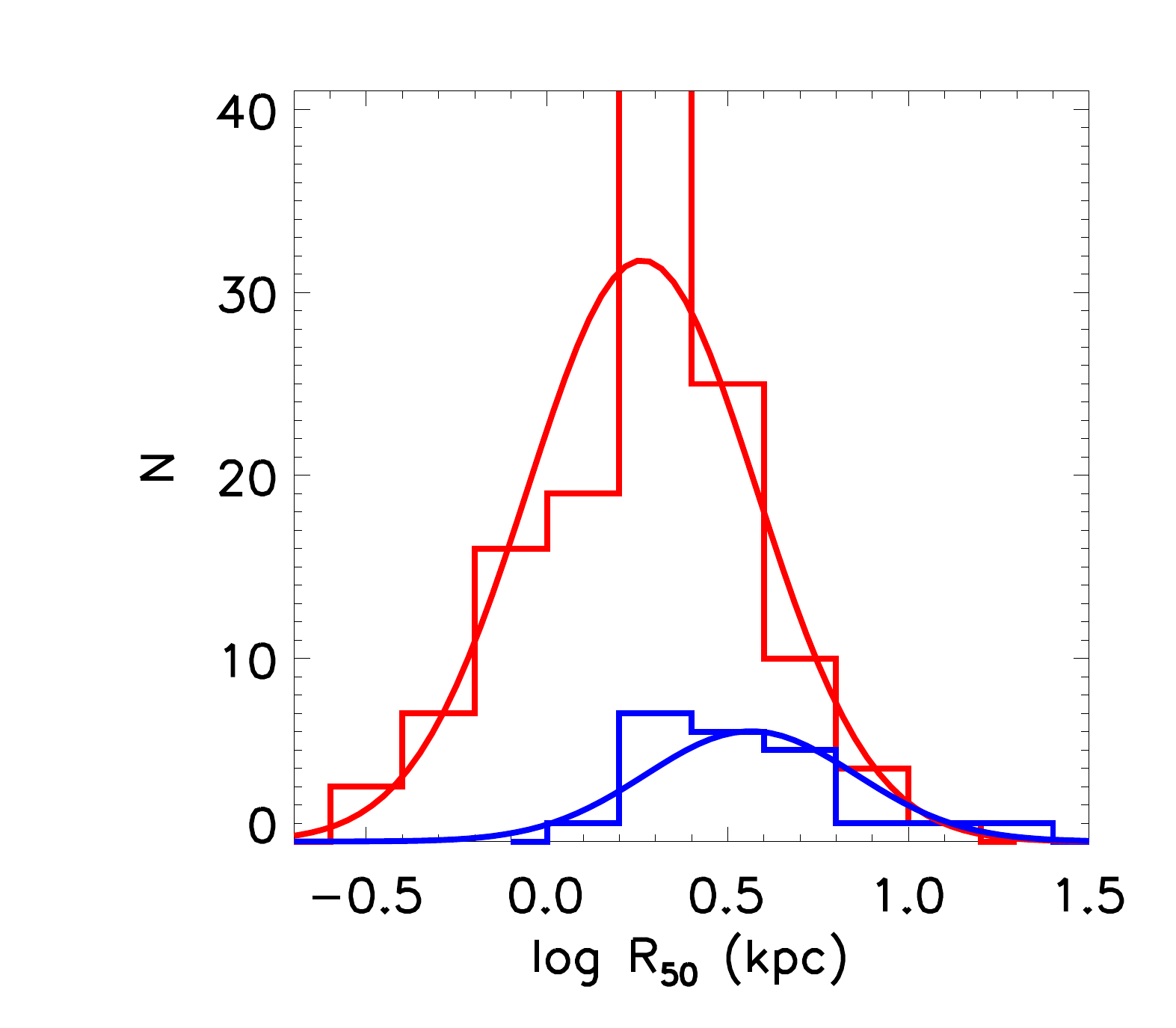}
\includegraphics[width=0.36\textwidth]{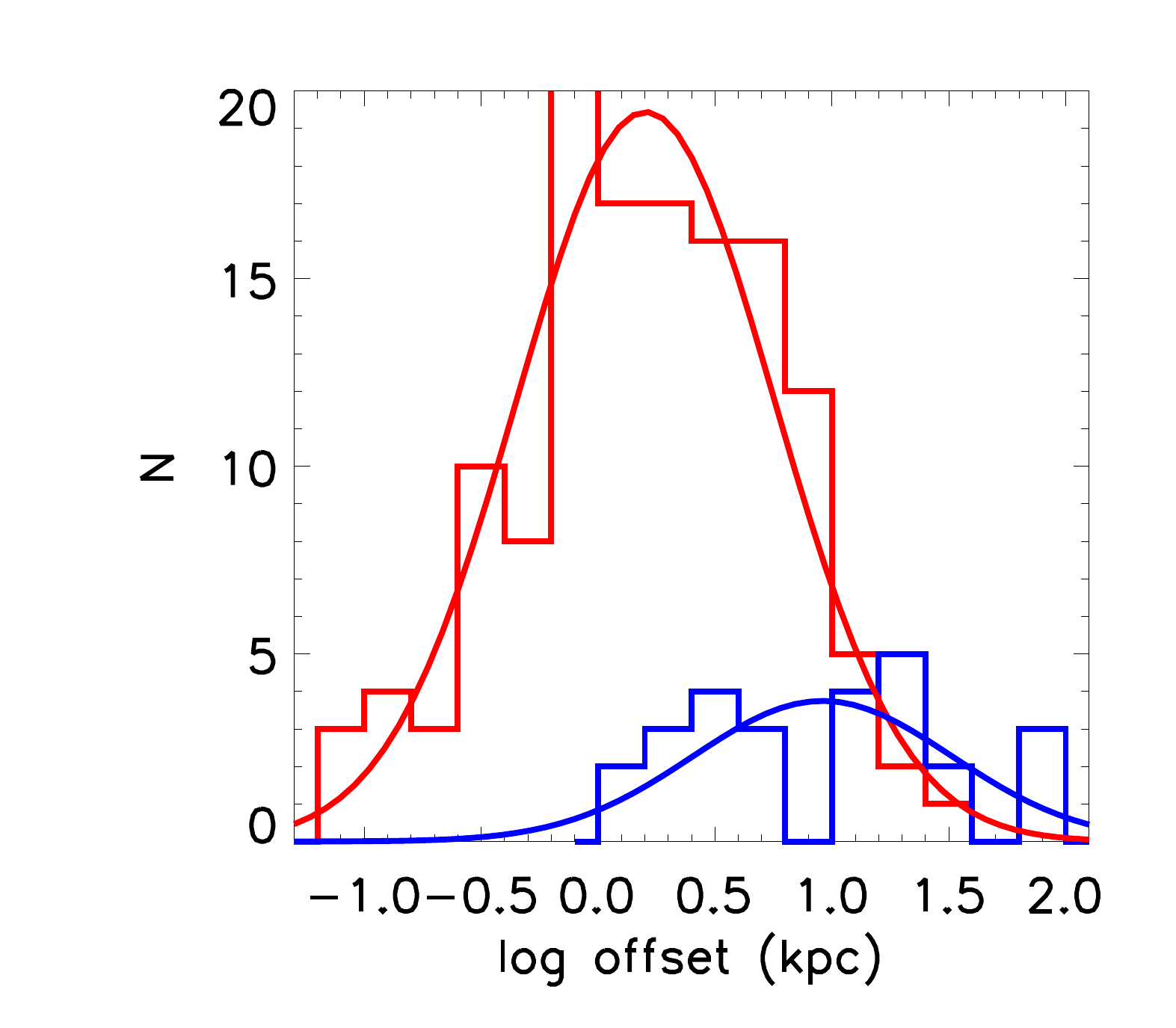}
\includegraphics[width=0.36\textwidth]{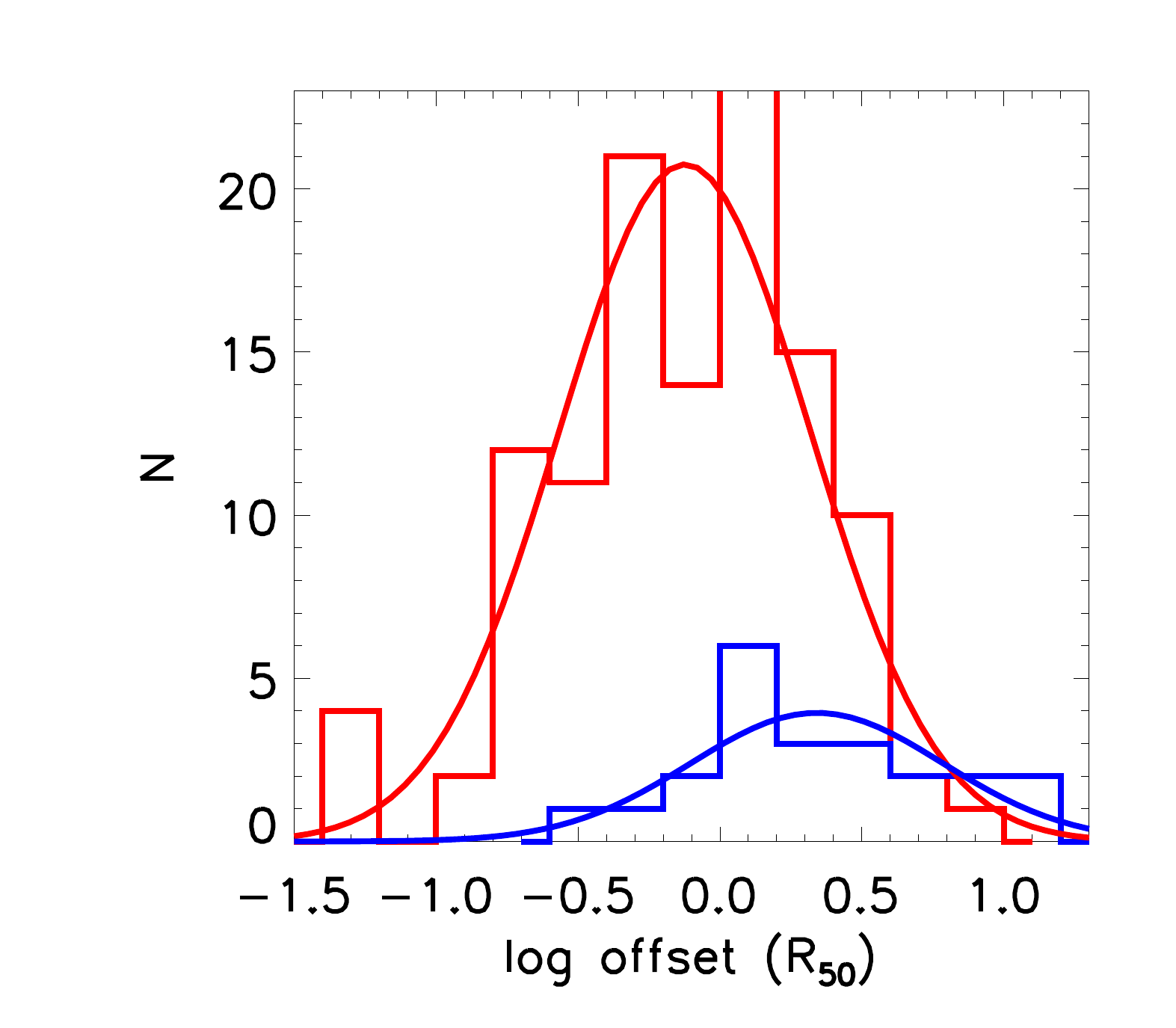}
\includegraphics[width=0.36\textwidth]{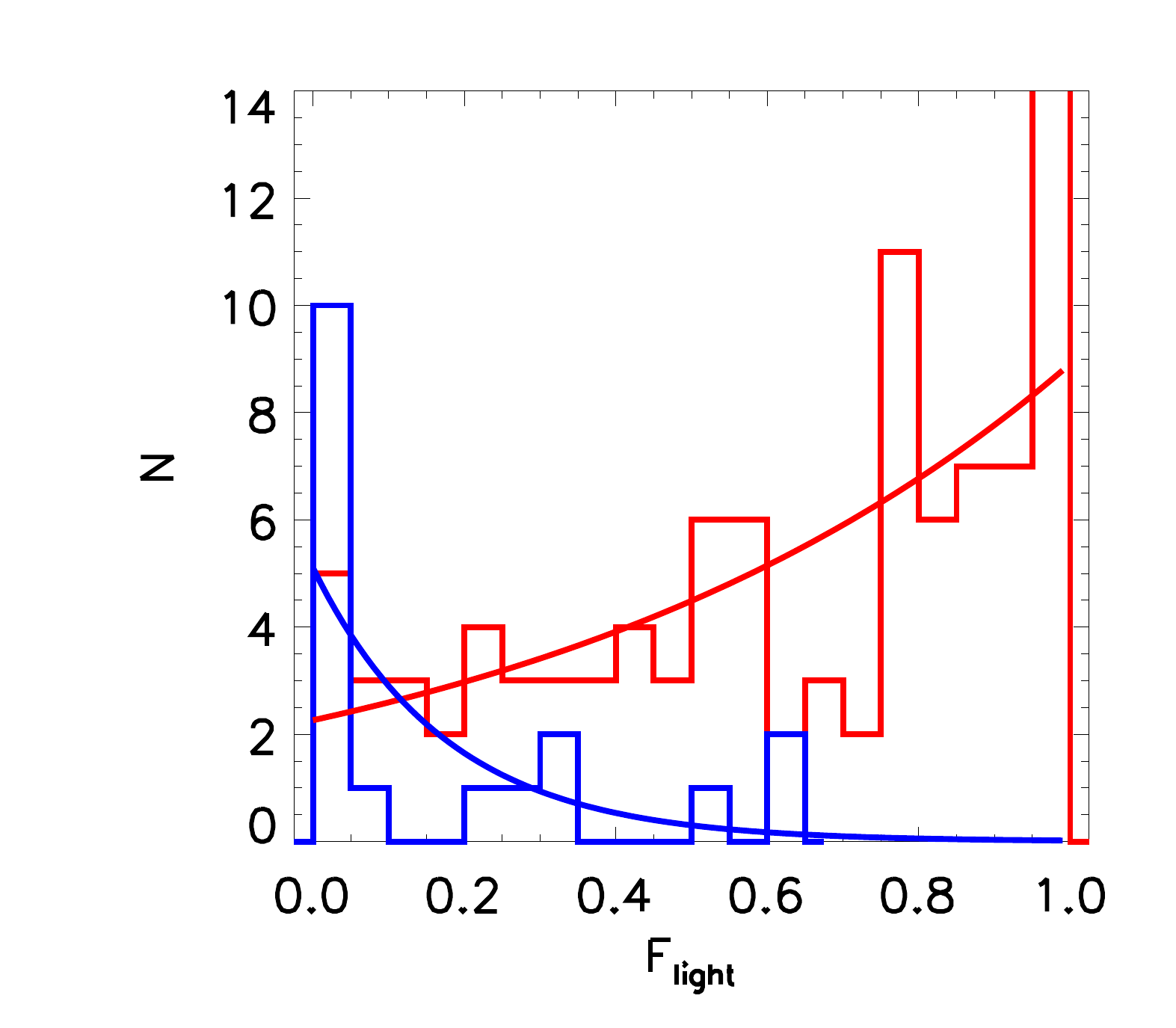}

\center{Fig. \ref{figfit}---Continued}
\end{figure*}

\begin{table*}[!htb] \scriptsize
\begin{center}
\caption{Fitting results of each parameter for the preliminary type II and I samples of \cite{2016ApJS..227....7L}. All parameters but $F_{\rm light}$ are fitted with a Gaussian distribution, and the distribution of $F_{\rm light}$ is fitted with a exponential distribution.}
\label{tbfit}
\begin{tabular}{l|ccccc|ccccc|c}
\hline
Parameter & \multicolumn{5}{c|}{Type II} & \multicolumn{5}{c}{Type I} \\
 & $N$ & mean $\mu$($\gamma$) & $\sigma$ & $D_{\rm KS}$ & $P_{\rm KS}$ & $N$ & mean$\mu$($\gamma$) & $\sigma$ & $D_{\rm KS}$ & $P_{\rm KS}$ & imputer$^3$ \\\hline
log $T_{90}$ & 371 & 1.68 $\pm$ 0.03 & 0.59 $\pm$ 0.03 & 0.05 & 0.41 & 32 & -0.36 $\pm$ 0.10 & 0.57 $\pm$ 0.08 & 0.07 & 1.00 & 0.65 \\
log (1+z) & 349 & 0.43 $\pm$ 0.01 & 0.18 $\pm$ 0.01 & 0.03 & 0.83 & 24 & 0.17 $\pm$ 0.02 & 0.08 $\pm$ 0.01 & 0.13 & 0.77 & 0.30 \\
log $E_{\rm iso}$ (erg) & 371 & 52.62 $\pm$ 0.05 & 1.01 $\pm$ 0.05 & 0.08 & 0.03 & 31 & 50.92 $\pm$ 0.16 & 0.94 $\pm$ 0.11 & 0.11 & 0.84 & 51.78 \\
log $L_{\rm iso}$ (erg s$^{-1}$) & 368 & 52.00 $\pm$ 0.06 & 1.13 $\pm$ 0.07 & 0.08 & 0.03 & 31 & 51.34 $\pm$ 0.19 & 1.07 $\pm$ 0.11 & 0.08 & 0.99 & 51.76 \\
$\alpha$ & 197 & -1.02 $\pm$ 0.03 & 0.34 $\pm$ 0.02 & 0.06 & 0.55 & 15 & -0.57 $\pm$ 0.06 & 0.26 $\pm$ 0.04 & 0.17 & 0.72 & -0.81 \\
log $E_{\rm p}$ (keV) & 203 & 2.16 $\pm$ 0.04 & 0.48 $\pm$ 0.03 & 0.05 & 0.69 & 17 & 2.63 $\pm$ 0.10 & 0.42 $\pm$ 0.08 & 0.14 & 0.89 & 2.36 \\
log $(f-1)$ & 222 & -0.37 $\pm$ 0.03 & 0.46 $\pm$ 0.02 & 0.09 & 0.04 & 28 & 0.17 $\pm$ 0.07 & 0.36 $\pm$ 0.03 & 0.11 & 0.84 & -0.14 \\
log ($f_{\rm eff}-1$) & 212 & -0.91 $\pm$ 0.02 & 0.34 $\pm$ 0.02 & 0.07 & 0.25 & 28 & 0.13 $\pm$ 0.07 & 0.37 $\pm$ 0.03 & 0.13 & 0.71 & -0.40 \\
log SFR (M$_{\odot}$ yr$^{-1}$) & 200 & 0.67 $\pm$ 0.05 & 0.82 $\pm$ 0.04 & 0.07 & 0.29 & 20 & 0.00 $\pm$ 0.19 & 0.87 $\pm$ 0.12 & 0.11 & 0.95 & 0.28 \\
log sSFR (Gyr$^{-1}$) & 92 & 0.06 $\pm$ 0.08 & 0.72 $\pm$ 0.06 & 0.08 & 0.56 & 15 & -0.53 $\pm$ 0.23 & 1.05 $\pm$ 0.15 & 0.12 & 0.98 & -0.57 \\
log $M_*$ ($M_{\odot}$) & 98 & 9.52 $\pm$ 0.08 & 0.81 $\pm$ 0.06 & 0.04 & 0.99 & 22 & 9.90 $\pm$ 0.18 & 0.84 $\pm$ 0.09 & 0.11 & 0.96 & 9.76 \\
$\rm [X/H]$ & 131 & -0.70 $\pm$ 0.06 & 0.62 $\pm$ 0.03 & 0.11 & 0.09 & 9 & -0.03 $\pm$ 0.05 & 0.16 $\pm$ 0.02 & 0.19 & 0.87 & -0.32 \\
log $R_{50}$ (kpc) & 126 & 0.26 $\pm$ 0.03 & 0.32 $\pm$ 0.02 & 0.07 & 0.50 & 22 & 0.56 $\pm$ 0.05 & 0.29 $\pm$ 0.05 & 0.13 & 0.81 & 0.39 \\
log offset (kpc) & 134 & 0.20 $\pm$ 0.05 & 0.55 $\pm$ 0.03 & 0.05 & 0.88 & 26 & 0.96 $\pm$ 0.11 & 0.55 $\pm$ 0.05 & 0.11 & 0.87 & 0.58 \\
log offset ($R_{50}$) & 115 & -0.12 $\pm$ 0.04 & 0.44 $\pm$ 0.03 & 0.06 & 0.75 & 22 & 0.34 $\pm$ 0.08 & 0.45 $\pm$ 0.06 & 0.12 & 0.88 & 0.11 \\
$F_{\rm light}$ & 97 & 1.37 $\pm$ 0.36$^1$ & - & 0.09 & 0.45 & 18 & -5.65 $\pm$ 1.88 & - & 0.56 & 0.00 & 0.36 \\
$F_{\rm light}$ & 97 & 1.33 $\pm$ 0.36$^2$ & - & 0.09 & 0.39 & 18 & -4.70 $\pm$ 1.88 & - & 0.25 & 0.19 & 0.38 \\
\hline 
\end{tabular}
\end{center}
\footnotesize
$^1$ For $F_{\rm light}$, this is the index $\gamma$ of the exponential distribution; \\
$^2$ After considering the rounding effect modification;\\ 
$^3$ The values to replace the missing values.
\end{table*}

\subsection{Classification \label{chp_classification}}

One of the most important problem in machine learning is how to deal with the missing values. Here we choose to impute the missing values with the equal-likelihood values, that is, the value to have the same likelihood to be Type I and Type II GRBs. The values are listed in the last column of Table \ref{tbfit}. 
It is equivalent to multiplying the likelihood $P(x_i|{\rm Type})$ with observed values only in Equation \ref{nb}, since $P(x_i|{\rm II})=P(x_i|{\rm I})$ for missing values.

``Naive'' Bayes assumes no correlation among parameters.
However, there are some obviously correlated parameters in our sample, e.g., the
subsets \{SFR, sSFR and M$_*$\}, \{$L_{\rm iso}$, $E_{\rm iso}$ and $T_{90}$\},
\{$R_{50}$, $R_{\rm off}$ and $r_{\rm off}$\}, and redshift $z$.
In order to eliminate significant correlations, 
{ we would like to exclude one parameter in each subset. Since we aim to distinguish the two types of GRBs, it is better to exclude the parameter with the least separations in each subset.  Table 7 of
\cite{2016ApJS..227....7L} tested the difference between each parameter of Type II and Type I by KS test, with $P_{\rm KS}$ indicating the difference between each parameter of Type II and Type I GRB. Thus, we exclude those parameters with the largest $P_{\rm KS}$, i.e., the least difference, in each parameter subset.}
For example, the isotropic peak
luminosity $L_{\rm iso}$ is excluded, since it could be roughly reproduced by
$T_{90}$ and $E_{\rm iso}$, and its $P_{\rm KS}$ value for Type II and Type I GRBs
is larger than those of $T_{90}$ and $E_{\rm iso}$.
Similarly, we exclude $f$, SFR, and $R_{\rm off}$. Finally, as presented in
\cite{2016ApJS..227....7L}, the redshift $z$ is subject to a selection effect
and is correlated with many parameters, such as $E_{\rm iso}$, SFR, and [X/H].
We therefore do not include $z$ in the analysis. In summary, we take the parameters 
$\{x\}=\{T_{90}, E_{\rm iso}, \alpha, E_{\rm p}, f_{\rm eff}, {\rm sSFR}, M_*, {\rm [X/H]}, R_{50},
r_{\rm off}=R_{\rm off}/R_{50}, F_{\rm light}\}$ to implement the Naive Bayes analysis.

The priors $P({\rm I})$ and $P({\rm II})$ are unknown.
We use the ratio of their observed numbers, i.e.,
$P({\rm II})=371/403=0.92$ and $P({\rm I})=32/403=0.08$
as the priors.


\section{Results}

\begin{table}[h]
    \centering
    \caption{A confusion matrix compares the preliminary GRB classification and the predicted classification of the Naive Bayes method described in Section 2.}
    \begin{tabular}{|c|cc|c|}
         \hline
         & Type II & Type I & total \\\hline
         Type II & 369 & 0 & 369 \\
         Type I & 2 &  32 & 34 \\\hline
         total & 371 & 32 & 403 \\
         \hline
    \end{tabular}
    \label{tab:results}
\end{table}

After { replacing (``imputing'')} the missing values with the equal-likelihood values listed in the last column of Table \ref{tbfit},
the posterior probability $P(\rm I|\{x\})$ and $P(\rm II|\{x\})$ can be calculated according to Equation \ref{nb}, with the likelihood estimated in Section \ref{chp_pdistribution} and the priors estimated in Section \ref{chp_classification}.
After normalizing the posterior probabilities 
$P(\rm I|\{x\})$
and $P(\rm II|\{x\})$, 
the type of each GRB is assigned to the one with a larger posterior probability.
We list the confusion matrix, which shows the number of false Type II (0), false Type I (2), true Type II (369), and true Type I (32), in Table \ref{tab:results}.
It can be shown that most GRBs are correctly classified as Type II or Type I, except 2 (0.5\%) GRBs.

\subsection{the Posterior Odds}
\begin{figure*}[!htb]
\centering
\includegraphics[width=0.48\textwidth]{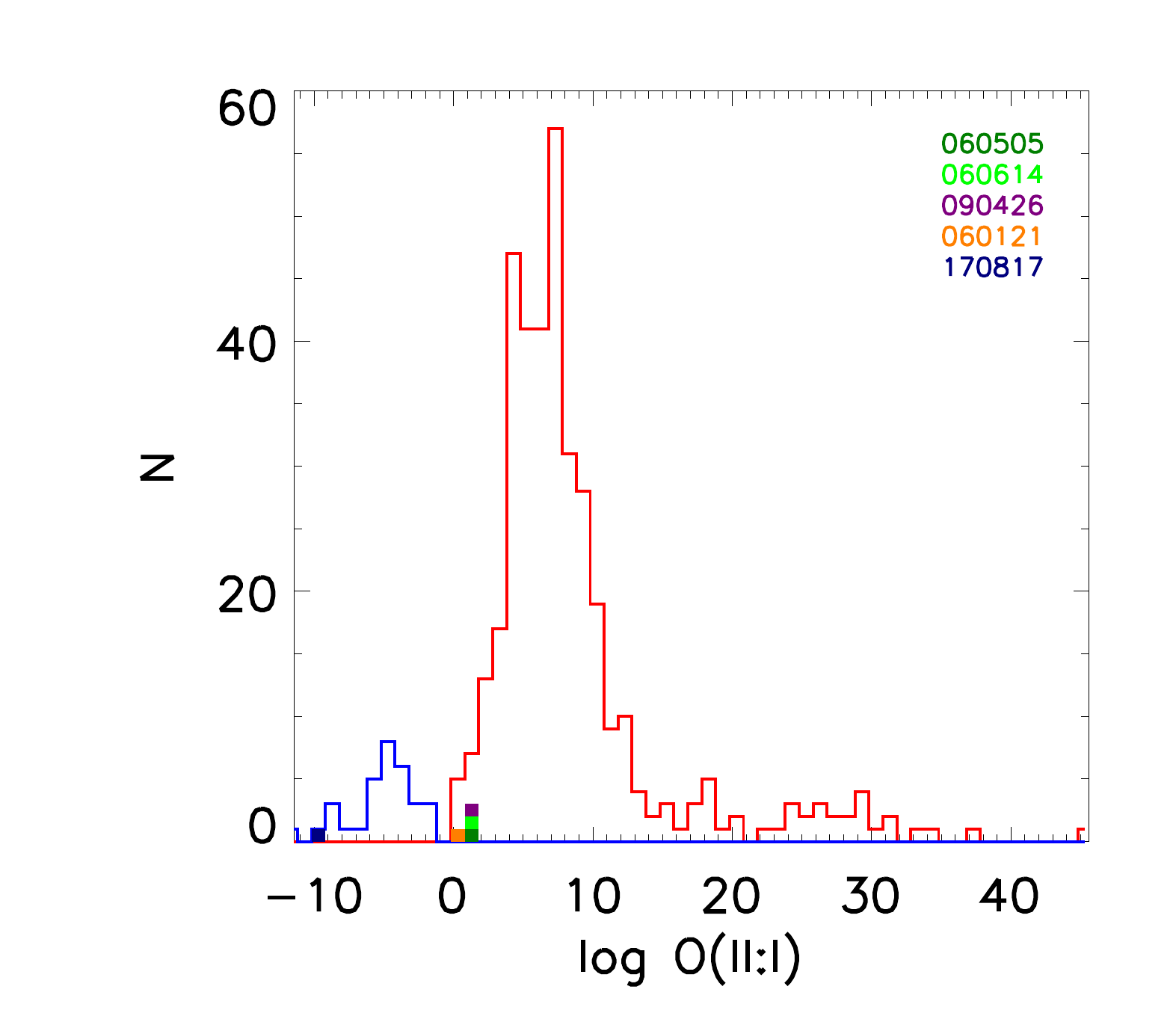}
\includegraphics[width=0.48\textwidth]{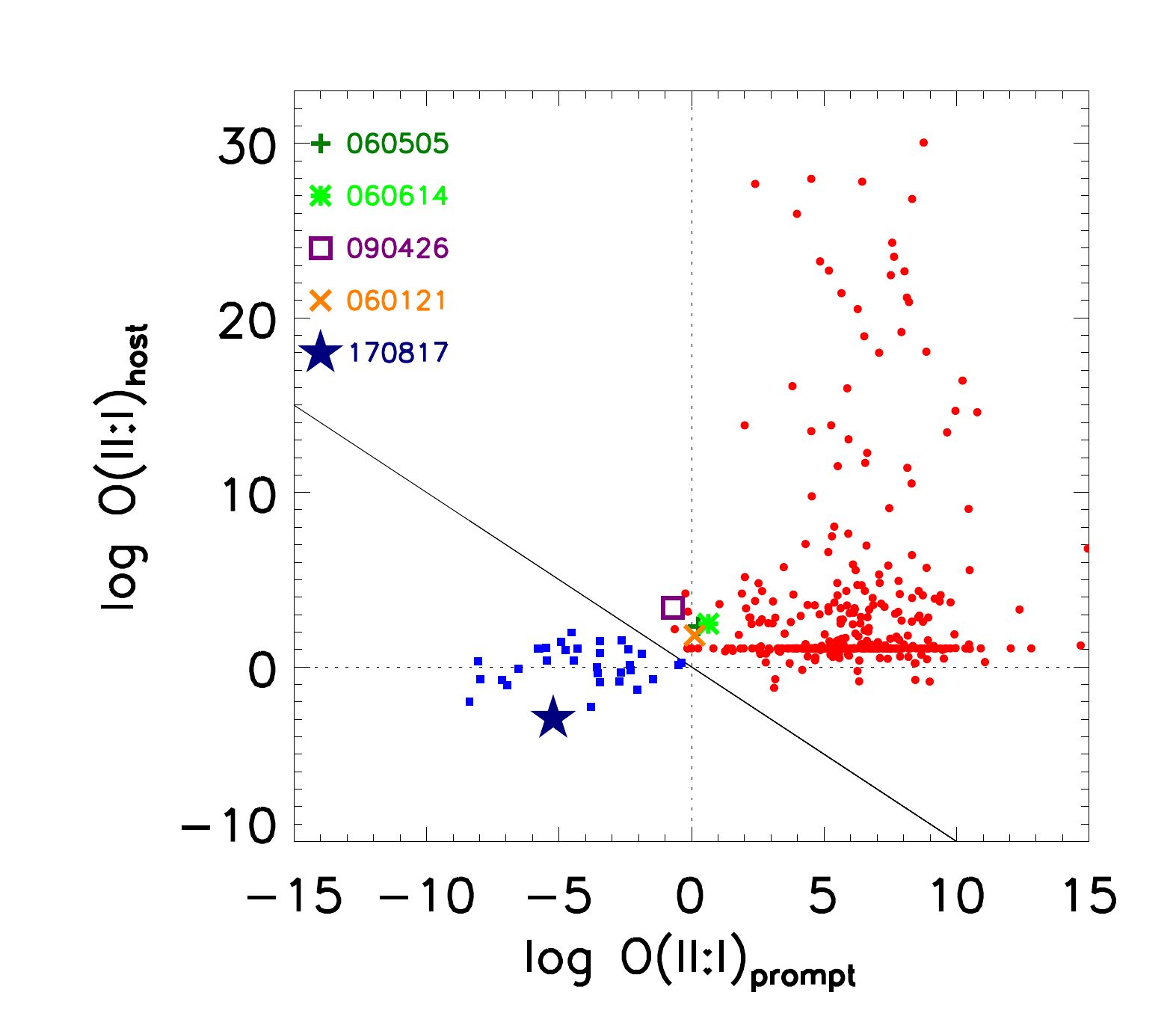}
\caption{
Left: The distribution of the posterior Odds log $O({\rm II:I}|\{x\})$. Red and blue histograms show the distribution of log $O({\rm II:I})$ for preliminary Type II GRBs (LGRBs) and Type I GRBs (SGRBs). 
The green and orange rectangles indicate the highly debated GRBs. The dark blue rectangle shows GRB 170817A, a ironclad Type I GRB associated with a binary neutron star merger. 
Right: The posterior Odds of the prompt emission properties log $O({\rm II:I})_{\rm prompt}$ and
host galaxy properties log $O({\rm II:I})_{\rm host}$.
Red dots indicate the preliminary Type II GRBs,
and blue squares indicate the preliminary Type I GRBs.
Green and orange symbols are the highly debated GRBs.
The dark blue star indicates GRB 170817A.
The solid line shows the separation line between preliminary Type I and preliminary Type II GRBs, log $O({\rm II:I})_{\rm host}=-{\rm log}\ O({\rm II:I})_{\rm prompt}$ line.
}
\label{fig:odds}
\end{figure*}

In order to examine the results in more detail, we calculate the posterior Odds 
\begin{eqnarray}
 O({\rm II:I}|\{x\})
 &=& \frac{P({\rm II}|\{x\})}{P({\rm I}|\{x\})} 
 = \frac{P({\rm II})}{P({\rm I})}
     \frac{P(\{x\}|{\rm II})}{P(\{x\}|{\rm I})},
\end{eqnarray}
which gives the degree that the observed parameter set $\{x\}$
supports Type II against Type I hypothesis.
The first term $P({\rm II})/P({\rm I})$ is the prior odds, and the second term $P(\{x\}|{\rm II})/P(\{x\}|{\rm I})$ is the Bayes factor (or likelihood ratio)
(\cite{odds} and their updates)\footnote{http://calcscience.uwe.ac.uk/Default.aspx}.

By definition, a positive log $O({\rm II:I})$ indicates a preference to Type II GRB, and a negative log $O({\rm II:I})$ indicates a preference to Type I. 

We plot the logarithmic posterior Odds log $O({\rm II:I})$ in the left panel of Fig. \ref{fig:odds}. 
Red and blue histograms show the distribution of log $O({\rm II:I})$ for preliminary Type II GRBs (LGRBs) and Type I GRBs (SGRBs). 
The logarithmic value $\log O({\rm II:I})$ of preliminary Type I GRBs are all negative, with the largest value $-1.2$. 
Most of preliminary Type II GRBs have positive log $O({\rm II:I})$, with two exceptions. 
They are GRB 131004A and GRB 090927. 
GRB 131004A has a duration $T_{90}=1.54 s$, and $f_{\rm eff}=1.94$, similar to a Type I GRB. However, the spectrum of it is quite soft, with a $\alpha=-1.4$ in Fermi/GBM detector, similar to a Type II GRB. 
And there is no more host galaxy information. Thus, the Naive Bayes classifier does not give a strong preference.
Its log $O({\rm II:I})=-0.16$, the smallest one of the preliminary Type II GRBs. 
GRB 090927A has $T_{90}=2.16 s$, $z=1.37$, { and  $\log O({\rm II:I})=-0.14$, showing no strong preference to either GRB type}.

The debated GRBs, GRB 060505, GRB 060614, GRB 090426, and GRB 060121 are not included in the histogram. They are presented as green and orange rectangles. They are all located near zero, consistent with the fact that their origins are in debate.
The ironic neutron star-neutron star merger event GRB 170817A is overplotted as the dark blue rectangle or star, which shows the smallest log $O({\rm II:I})$, indicating that it is a prototype of Type I GRBs.

Note that,  there is no overlap between the two preliminary samples of Type I and Type II GRBs.
The gap is between -1.2 and -0.16. 
It means that an additional modification \begin{equation}
{\cal  O}={\rm log}\ O({\rm II:I})+0.7
\label{eq:separation}
\end{equation}
would be a better criterion to classify GRBs physically. With it, every predicted type matches the preliminary type.

\subsection{Prompt emission versus host properties}

In order to examine the effect of prompt emission and host galaxy properties, we further study them separately.
The posterior Odds are
\begin{eqnarray*}
{\rm log}\ O({\rm II:I})_{\rm prompt} & = &
{\rm log}\ O({\rm II:I}|\{x\})_{\rm prompt}\\
& = & {\rm log}\ O({\rm II:I}|\{T_{90}, E_{\rm iso}, \alpha, E_{\rm peak}, f_{\rm eff}\})
\end{eqnarray*}
and
\begin{eqnarray*}
{\rm log}\ O({\rm II:I})_{\rm host} &=&
{\rm log}\ O({\rm II:I}|\{x\})_{\rm host} \\
&=&{\rm log}\ O({\rm II:I}|\{{\rm sSFR}, M_*, {\rm [X/H]}, R_{50}, r_{\rm off}, F_{\rm light}\}).
\end{eqnarray*}
The results are presented in the right panel of Fig. \ref{fig:odds}.
The preliminary Type II GRBs are shown with red points, and the preliminary Type I
GRBs are shown with blue squares. 
{ The GRBs without host galaxy information have ${\rm log}\ O({\rm II:I})_{\rm host}={\rm log}\ P({\rm II})/P({\rm I})=1.06$, and cluster in the figure. }
{ By definition}, a positive log $O({\rm II:I})_{\rm prompt}$
or log $O({\rm II:I})_{\rm host}$ indicates the preference of a Type II GRB over
a Type I GRB. It can be seen that most Type II GRB candidates (red dots) are indeed
located in the first quadrant, with log $O({\rm II:I})_{\rm prompt} > 0$ and
log $O({\rm II:I})_{\rm host} > 0$, and { lots of} Type I candidates (blue dots) are
located in the third quadrant. 
Also, there are a few preliminary Type II GRBs in the second and forth quadrant, 
and there are some preliminary Type I GRBs in the second quadrant.
It may be results of the correlations among parameters, or the { simple estimation of the priors by the detected Type II/Type I GRBs}.
Still, in the ${\rm log}\ O({\rm II:I})_{\rm host}$ vs ${\rm log}\ O({\rm II:I})_{\rm prompt}$ diagram, we find that a straight line
\begin{equation}
{\rm log}\ O({\rm II:I})_{\rm host} = - {\rm log}\ O({\rm II:I})_{\rm prompt},
\end{equation}
is able to separate these two types of GRBs, as shown by the solid line.

All the four debated GRBs
are quite close to the Type I region in the right panel of Fig.\ref{fig:odds}, although they are located in the Type II
region. GRB 060121 has a $T_{90}$ of 2.61 s, longer than the conventional separation line for SGRBs and LGRBs.
The low energy photon index $\alpha$ is $-0.5$, around the typical value of Type I GRBs.
The size of its host is $R_{50}=10^{0.6}$ kpc, which is also a typical value of Type
I GRBs. On the other hand, its offset and fraction of light $F_{\rm light}$ suggest that it belongs to Type II.
This object is somewhat between Type II and Type I GRBs. Note that there is no
convincing redshift measurement of GRB 060121A, and a redshift of $z=0.5$ has been assumed \citep{2016ApJS..227....7L}.

GRB 170817A, the only ironclad Type I GRB firmly associated with a binary neutron star merger from the gravitational wave observations,
is presented as a dark blue star.
The prompt emission properties are taken from \cite{2018NatCo...9..447Z},
and the host galaxy properties are taken from \citep{2017ApJ...848L..22B}.
Its log $O$(II:I)$_{\rm prompt}$ and log $O$(II:I)$_{\rm host}$ are -5.2 and -3.8, respectively.
Both are well below the separation line, well consistent with our Type I GRB classification criterion.

\subsection{SN and SN limits}

\begin{figure*}[!htb]
\centering
\includegraphics[width=0.48\textwidth]{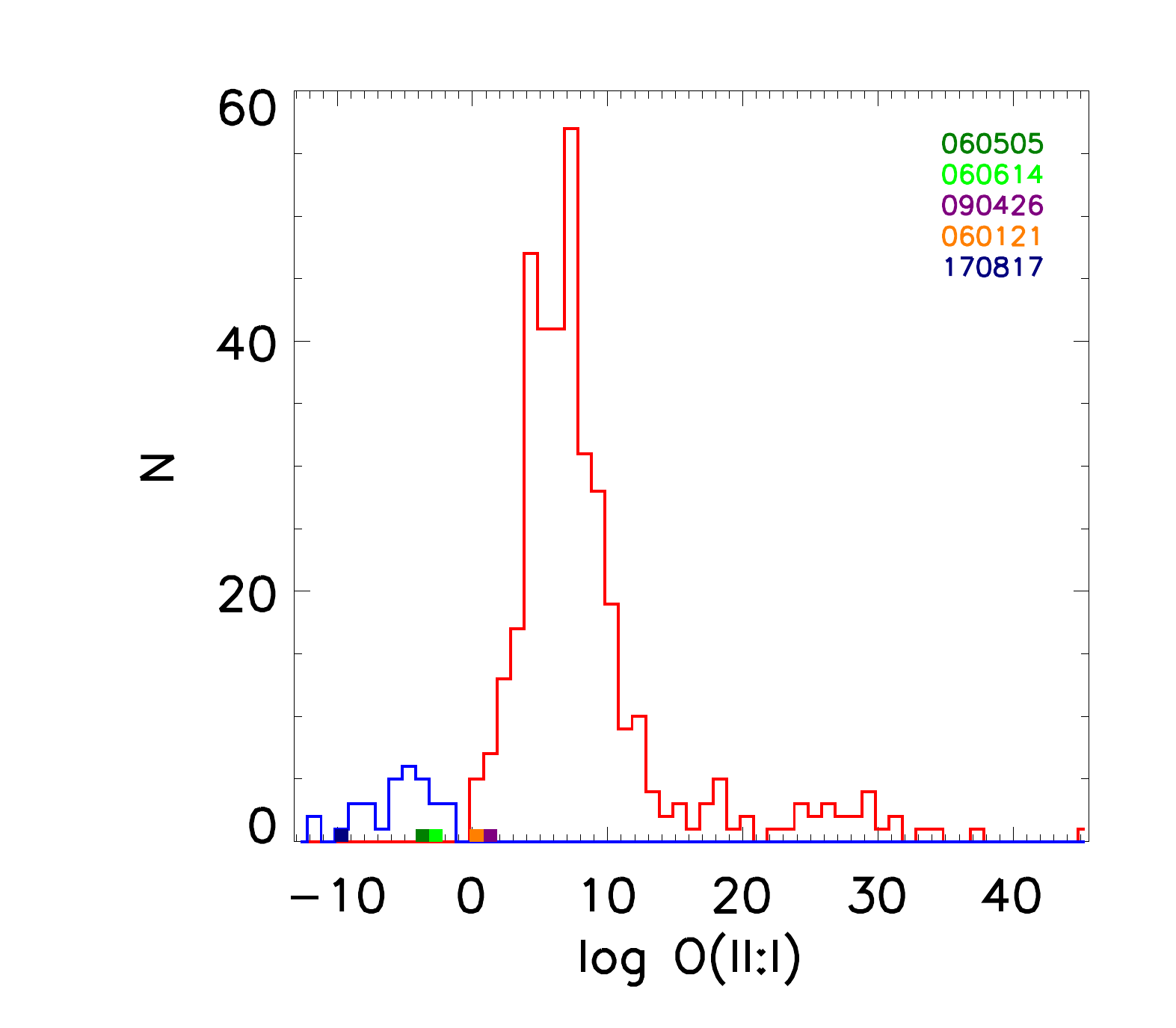}
\includegraphics[width=0.48\textwidth]{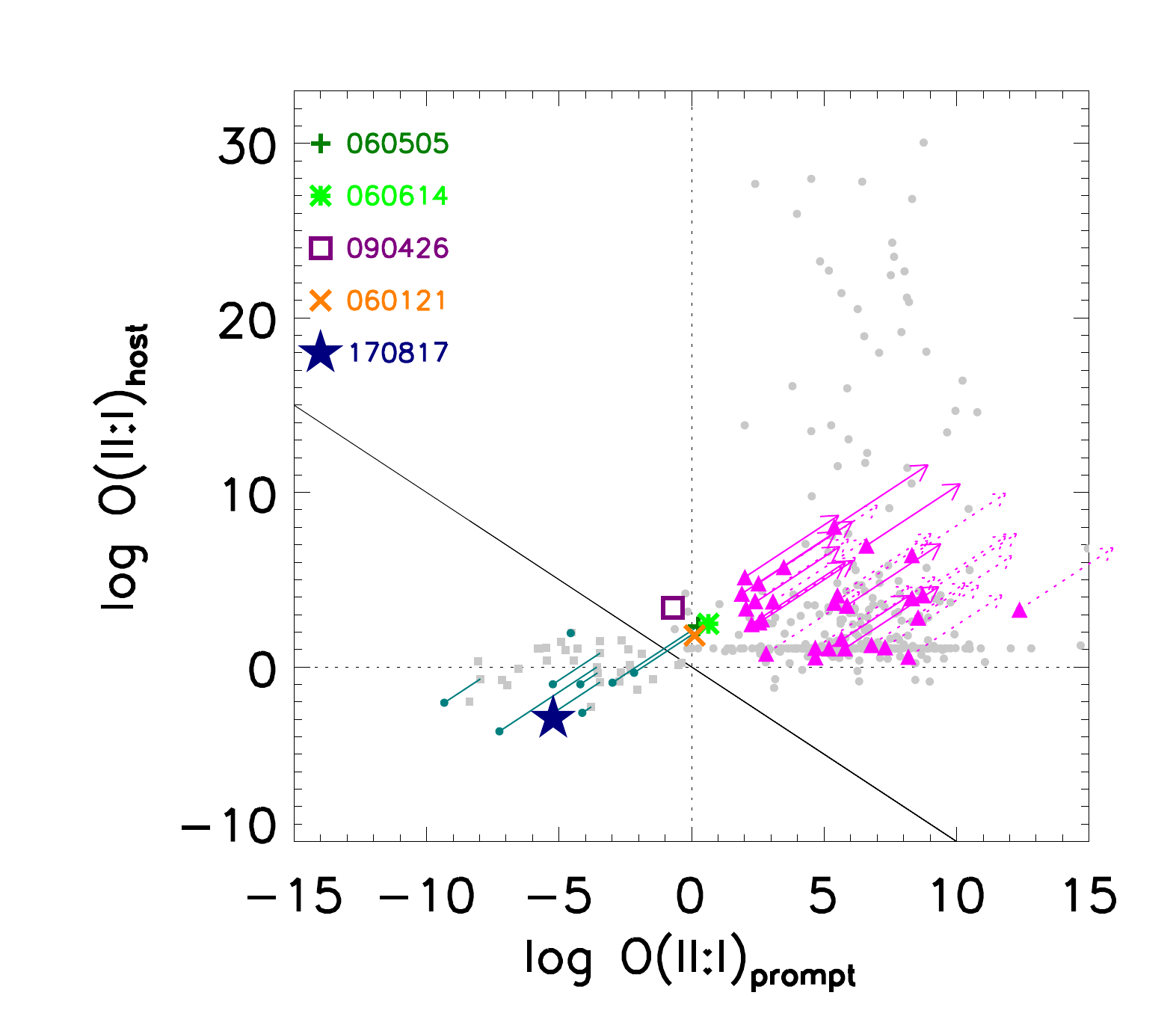}
\caption{
Same as Figure \ref{fig:odds}, but with the supernovae information added.
Right: Triangles represent the GRBs with SN associations, with spectral (solid arrows) or
photometric (dashed arrows) detections. Dark cyan lines and points show the correction
with SN limits included.
The GRBs without SN information are plotted
as gray for clearness.
}
\label{fig:sn}
\end{figure*}

The SN association and observational limits of the existence of SN provide important information about the physical origin of GRBs.
{ We add the SN information to posterior Odds calculations and they are presented in Fig. \ref{fig:sn}.}

For some GRBs, stringent limits of SNe were set, which give a strong
indication of their origins. The inclusion of SN limits leads to a correction of
the posterior Odds.
{ Assuming a Gaussian distribution of the peak absolute magnitudes of the associated SNe, we estimate the probability to have an associated SN fainter than the upper limit $P({\rm SN})$ with the Gaussian distribution and modify the Odds to be log $O_{\rm sn}({\rm II:I})={\rm log}\ O({\rm II:I})+{\rm log}\ P({\rm SN})$. 
The peak absolute magnitudes of the observed GRB-associated SNe have a narrow distribution, $-18.52\pm0.45$ mag \citep{2014ARA&A..52...43B, 2012grbu.book..169H}. However, it is possible that fainter associated SNe are not detected. 
Since most of the LGRB-associated SNe are Type Ic with broad lines, we adopt the peak absolute magnitude distribution of SN Ic-BL as the distribution of that of LGRB-SNe. It is estimated to be $-18.3\pm1.6$ mag from the Open SN catalog\footnote{https://sne.space/}.
The results of GRB classifications after including the SN criterion are presented in Figure \ref{fig:sn}. 
Because the GRBs nearest to the gaps usually do not have SN information, the position of the gap between Type I and Type does not change. 
}

With the SN limit correction,
two of the highly debated object, Type II candidates, GRB 060614 and 060505, are now shifted to the Type I
region. These two GRBs are also widely discussed to be long-duration Type I candidates
\citep{2006Natur.444.1044G,2006Natur.444.1050D, 2006Natur.444.1047F,2007ApJ...655L..25Z}.
Their prompt emission properties are not that typical compared with other Type II GRBs,
and their host galaxy properties are similar to Type II GRBs. However, their SN limits { strongly favor the merger origin of these two events}.

The SN limits are also presented in the log $O({\rm II:I})_{\rm host}$-log $O({\rm II:I})_{\rm prompt}$ diagram (the right panel of Fig. \ref{fig:sn}), as dark cyan lines.

GRBs with spectrally confirmed SN associations are definitely Type II GRBs from core-collapse of massive stars. 
In the right panel of Fig. \ref{fig:sn}, GRBs with spectrally confirmed SN associations are shown as triangles with solid arrows, and those with SN spectral features are presented as triangles with dashed arrows
\citep{2012grbu.book..169H}. 

\section{Conclusion and Discussion}

Utilizing the distributions of the prompt emission and host galaxy
properties of GRBs presented in our previous work \citep{2016ApJS..227....7L}, in this paper
we proposed to {\rm use Naive Bayes method} to classify GRBs into two physically distinct categories, Type II (massive star origin) and Type I GRBs (compact star origin). 
We estimate the probability of each GRB to be a Type II or Type I GRB
based on the distributions of the prompt emission and host galaxy parameters derived from the preliminary Type II and Type I GRB samples. 
{The type of each GRB is assigned to the one with the larger probability.}
This results in only 0.5\% mis-classification, much better than any one-parameter criteria.

We also examined the posterior odds, log $O({\rm II:I})$, which is the logarithm
of the ratio of the probabilities of Type II and Type I, and describes the preference
of Type II against Type I. 
{ For the GRB control sample (described in Section 2.1)}, the two preliminary Type I and Type II
samples are well separated into two groups without an overlap in the log $O({\rm II:I})$ distribution, with a gap between -1.2 and -0.16. 
{ If we define Eq.(\ref{eq:separation}),
Type I and Type II GRBs can be classified as ${\cal O} < 0$ and ${\cal O} > 0$, respectively.
Such a new classification scheme is more  efficient, resulting in no mis-classification relative to the preliminary sample.}
We believe that this method provides a quantitative, overall
assessment of the physical origin of any GRB with multi-wavelength observational data in the future.
{ We provide an interface website and a public python code in \href{http://www.physics.unlv.edu/~liye/GRB/grb_cls.html}{http://www.physics.unlv.edu/$\sim$liye/GRB/grb\_cls.html}, which can be directly used to classify future GRBs with desired observational information available.}

We discuss some caveats and possible improvements of this method in the following.

\subsection{Validation}

{ Machine learning studies reveal that the errors in the training sample, the sample to estimate the parameter distributions, are usually smaller than the true error. 
A common method to estimate the true error is cross validation. By randomly dividing the total sample into a training sample and a test sample, which does not contribute to the parameter estimation, the true error is around the error of the test sample. 
However, the sample size of the preliminary Type I GRBs is only 32. An equal division would significantly reduce the sample size, and overestimate the test error.
In order to eliminate this issue, 
we utilize the ``Leave One Out Cross Validation (LOOCV)''. 
This method runs the Naive Bayes method 403 times, since we have 403 GRBs in the controlled sample.
For each time, one GRB is left out as a test sample, while the other 402 GRBs are used as a training sample to fit the distributions of the parameters. 
The sum of the errors of the 403 realizations is considered as the test error.
As a result, 4 Type II GRBs are misclassified as Type I GRBs, and no Type I GRBs are misclassified as Type II GRBs.
The test error rate is thus 4/403=1\%. 
}

\subsection{Selection of the control sample}

The selection of the control sample might affect our results. We, therefore, test
the case with the four highly debated GRBs in the analysis. 
{ With well-defined the Naive Bayes method, there are 3 preliminary Type II GRBs mis-classifed as Type I GRBs, and 2 preliminary Type I GRB mis-classified as Type II GRBs. 
The 2 mis-classified Type I GRBs are GRB 090426 and GRB 060121, two of the highly debated GRBs.
With the log $O'=$ log $O({\rm II:I})+0.7$ as a criteria, only the two highly debated GRB 090426 and GRB 060121 are mis-classified.
}

There is a slight overlap between the two candidate populations. The
overlap fraction is { 2\% (8)} for Type II candidates and { 6\% (2)} for Type I candidates.
Even in this case, the overlap fraction is much smaller than 
previous methods. For example, for the duration classification method, there are 7\% Type
II and 20\% Type I candidates in the overlapping region. For the $f_{\rm eff}$ parameter
method, the corresponding fraction is 7\% (48\%) for Type II (Type I) candidates, respectively.

\subsection{Selection of parameters}
{ 
The Naive Bayes method assumes no correlation among parameters. 
To eliminate strong correlations, 
we select to use $\{x\}= \{T_{90},
E_{iso}, \alpha, E_{\rm p}, f_{\rm eff}, {\rm sSFR}, M_*, {\rm [X/H]}, R_{50}, r_{\rm off}, F_{\rm light} \}$ only. 
Since Naive Bayes performs amazingly well in mildly correlated cases,
we also wonder what would be the result if we use different parameter groups.

In order to test this, we include all the 16 available parameters in Table \ref{tbfit}.
It turns out that there are 5 preliminary Type II GRBs mis-classifed as Type I GRBs, and no preliminary Type I GRB mis-classified as Type II GRBs. 
There is a small overlap region between [-2.84, -2.68], which covers one preliminary Type II GRB and one preliminary Type I GRB. 
It is clear that, even with the strong correlated parameters, Naive Bayes method still performs better than the one-parameter criteria, although not as good as what we used in Section 2. 
}

\subsection{Selection of functions}
{
In Section \ref{chp_pdistribution}, we use a Gaussian function to fit the parameter distribution of each GRB class except $F_{\rm light}$. The fitting results of all parameters are acceptable with a significance level 0.01, when the goodness of fit are examined with Kolmogorov-Smirnov (KS) test. 

{ On the other hand, Anderson-Darling (AD) test is usually considered as a better proxy of Gaussianity, especially the tails. }When AD test is applied to the examination, the null probability for $E_{\rm iso}$, $L_{\rm peak}$, $f$, and [X/H] of Type II GRB are smaller than 0.01. 
{ It indicates that they may not be well described as Gaussian distributions.}
For $E_{\rm iso}$ and $L_{\rm peak}$, the low-luminosity GRBs contribute as a low-luminosity tail  \citep{2007ApJ...662.1111L}. For [X/H], it is a result of different metallicity estimation methods or the metallicity dependence on redshift. For low redshift objects, the metallicities are estimated with emission lines such as $H\alpha$, OII, while for high redshift objects, they are mainly estimated with absorption lines. 
We thus tried to use two Gaussians for the above four parameters of Type II GRB. With two Gaussians, the resulting AD test parameter $A^2$(and null probabilities $P$) are 0.38(0.56), 0.98(0.14), 0.29(0.69) and 0.26(0.70), respectively.
The classification results are nearly the same. One more preliminary Type II GRB, GRB 070714A, has a log O smaller than 0. The gap between preliminary Type II and Type I GRBs still exists, just change to [-1.23, -0.45].

Theoretically, AD test is more sensitive to the tail, KS test is more sensitive to the center, and Cramer-von Mises test is somewhat in between. The additional Gaussian contributes 0.05, 0.10, 0.27 and 0.19 to $E_{\rm iso}$, $L_{\rm peak}$, $f$ and [X/H] respectively, while change the Anderson-Darling test results much.
The choice of goodness of fit methods depends on the aim of the test. 
For our purpose, the centers of Type I and Type II GRBs are more important than the tails. 
Since the results do not change significantly,
to have a clear and consistent picture, we use one Gaussian in the main part of the paper. 
Two-Gaussian fitting reveal more details and may be needed if more detailed classifications, including subclasses, is to be operated. 
}

\subsection{Other possible physical classes/subclasses of GRBs}

{ The purpose of physical classification of GRBs is to identify physically distinct classes using (typically multiple) observational criteria. The most important physical question in GRB physics is the progenitor systems. Our Type I/II classification scheme developed in this paper (proposed in \cite{2009ApJ...703.1696Z}) is dedicated to address this problem.}
Notice that such a classification scheme does not specify the detailed progenitor system.
While the leading progenitor model of Type II GRBs is collapse of single stars (collapsars)
\citep{1993ApJ...405..273W, 1999ApJ...524..262M}, 
these GRBs may be also produced by massive stars in binary systems \citep[e.g.][]{2004MNRAS.348.1215I, 2005ApJ...623..302F}. 
The leading Type I progenitor model is NS-NS mergers
\citep{1986ApJ...308L..43P, 1989Natur.340..126E, 1992ApJ...395L..83N}, 
but BH-NS mergers are quite possible to be another type of progenitor systems \citep{1991AcA....41..257P, 2006ApJ...641L..93F}. 
{ It is possible that there exist sub-classes within each broad class (e.g. \cite{Zhang2018book} for a detailed discussion). For example, 
within the Type I population, those with extended emission or internal plateau may comprise a sub-class of NS-NS merger events that have a rapidly spinning neutron star post-merger product
\citep{2001A&A...379L..39L, 2006ApJ...643..266N, 2010ApJ...717..411N, 2015MNRAS.452..824K}. 
Within the Type II category, the so-called low-luminosity GRBs \citep{2007ApJ...662.1111L} and ultra-long GRBs \citep{2013ApJ...766...30G,2014ApJ...781...13L,2014ApJ...787...66Z} may signify different sub-classes. Also, if one is interested in radiation mechanisms and jet compositions, GRBs may be classified as thermally-dominated fireballs and Poynting-flux-dominated jets.}
It is possible to further identify physically-distinct sub-classes or even new classes with the multi-wavelength data. This is beyond the scope of the current paper.

\acknowledgments
{ We thank the referee for helpful comments and suggestions.}
YL thank Seng, Fei and Jun Qin for helpful discussion.
This work is partially supported by the China Postdoctoral Science Foundation (No. 2018M631242).
YL is supported by the KIAA-CAS Fellowship, which is jointly supported by Peking University and Chinese Academy of Sciences. 
QY acknowledges the support of the 100 Talents Program of Chinese Academy of Sciences.


\begin{thebibliography}{}
\expandafter\ifx\csname natexlab\endcsname\relax\def\natexlab#1{#1}\fi

\bibitem[{{Abbott} {et~al.}(2017{\natexlab{a}}){Abbott}, {Abbott}, {Abbott},
  {Acernese}, {Ackley}, {Adams}, {Adams}, {Addesso}, {Adhikari}, {Adya}, \&
  et~al.}]{2017ApJ...848L..13A}
{Abbott}, B.~P., {Abbott}, R., {Abbott}, T.~D., {et~al.} 2017{\natexlab{a}},
  \apjl, 848, L13

\bibitem[{{Abbott} {et~al.}(2017{\natexlab{b}}){Abbott}, {Abbott}, {Abbott},
  {Acernese}, {Ackley}, {Adams}, {Adams}, {Addesso}, {Adhikari}, {Adya}, \&
  et~al.}]{2017PhRvL.119p1101A}
---. 2017{\natexlab{b}}, Physical Review Letters, 119, 161101

\bibitem[{{Abbott} {et~al.}(2017{\natexlab{c}}){Abbott}, {Abbott}, {Abbott},
  {Acernese}, {Ackley}, {Adams}, {Adams}, {Addesso}, {Adhikari}, {Adya}, \&
  et~al.}]{2017ApJ...848L..12A}
---. 2017{\natexlab{c}}, \apjl, 848, L12

\bibitem[{{Antonelli} {et~al.}(2009){Antonelli}, {D'Avanzo}, {Perna}, {Amati},
  {Covino}, {Cutini}, {D'Elia}, {Gallozzi}, {Grazian}, {Palazzi},
  {Piranomonte}, {Rossi}, {Spiro}, {Stella}, {Testa}, {Chincarini}, {di Paola},
  {Fiore}, {Fugazza}, {Giallongo}, {Maiorano}, {Masetti}, {Pedichini},
  {Salvaterra}, {Tagliaferri}, \& {Vergani}}]{antonelli2009}
{Antonelli}, L.~A., {D'Avanzo}, P., {Perna}, R., {et~al.} 2009, \aap, 507, L45

\bibitem[{{Band} {et~al.}(1993){Band}, {Matteson}, {Ford}, {Schaefer},
  {Palmer}, {Teegarden}, {Cline}, {Briggs}, {Paciesas}, {Pendleton}, {Fishman},
  {Kouveliotou}, {Meegan}, {Wilson}, \& {Lestrade}}]{1993ApJ...413..281B}
{Band}, D., {Matteson}, J., {Ford}, L., {et~al.} 1993, \apj, 413, 281

\bibitem[{{Berger}(2009)}]{2009ApJ...690..231B}
{Berger}, E. 2009, \apj, 690, 231

\bibitem[{{Berger}(2014)}]{2014ARA&A..52...43B}
---. 2014, \araa, 52, 43

\bibitem[{{Berger} {et~al.}(2013){Berger}, {Fong}, \&
  {Chornock}}]{2013ApJ...774L..23B}
{Berger}, E., {Fong}, W., \& {Chornock}, R. 2013, \apjl, 774, L23

\bibitem[{{Berger} {et~al.}(2005){Berger}, {Price}, {Cenko}, {Gal-Yam},
  {Soderberg}, {Kasliwal}, {Leonard}, {Cameron}, {Frail}, {Kulkarni}, {Murphy},
  {Krzeminski}, {Piran}, {Lee}, {Roth}, {Moon}, {Fox}, {Harrison}, {Persson},
  {Schmidt}, {Penprase}, {Rich}, {Peterson}, \& {Cowie}}]{2005Natur.438..988B}
{Berger}, E., {Price}, P.~A., {Cenko}, S.~B., {et~al.} 2005, \nat, 438, 988

\bibitem[{{Blanchard} {et~al.}(2016){Blanchard}, {Berger}, \&
  {Fong}}]{2016ApJ...817..144B}
{Blanchard}, P.~K., {Berger}, E., \& {Fong}, W.-f. 2016, \apj, 817, 144

\bibitem[{{Blanchard} {et~al.}(2017){Blanchard}, {Berger}, {Fong}, {Nicholl},
  {Leja}, {Conroy}, {Alexander}, {Margutti}, {Williams}, {Doctor}, {Chornock},
  {Villar}, {Cowperthwaite}, {Annis}, {Brout}, {Brown}, {Chen}, {Eftekhari},
  {Frieman}, {Holz}, {Metzger}, {Rest}, {Sako}, \&
  {Soares-Santos}}]{2017ApJ...848L..22B}
{Blanchard}, P.~K., {Berger}, E., {Fong}, W., {et~al.} 2017, \apjl, 848, L22

\bibitem[{{Bloom} {et~al.}(1998){Bloom}, {Djorgovski}, {Kulkarni}, \&
  {Frail}}]{1998ApJ...507L..25B}
{Bloom}, J.~S., {Djorgovski}, S.~G., {Kulkarni}, S.~R., \& {Frail}, D.~A. 1998,
  \apjl, 507, L25

\bibitem[{{Bloom} {et~al.}(2002){Bloom}, {Kulkarni}, \&
  {Djorgovski}}]{2002AJ....123.1111B}
{Bloom}, J.~S., {Kulkarni}, S.~R., \& {Djorgovski}, S.~G. 2002, \aj, 123, 1111

\bibitem[{{Broos} {et~al.}(2011){Broos}, {Getman}, {Povich}, {Townsley},
  {Feigelson}, \& {Garmire}}]{2011ApJS..194....4B}
{Broos}, P.~S., {Getman}, K.~V., {Povich}, M.~S., {et~al.} 2011, \apjs, 194, 4

\bibitem[{{Broos} {et~al.}(2013){Broos}, {Getman}, {Povich}, {Feigelson},
  {Townsley}, {Naylor}, {Kuhn}, {King}, \& {Busk}}]{2013ApJS..209...32B}
---. 2013, \apjs, 209, 32

\bibitem[{{Chary} {et~al.}(2002){Chary}, {Becklin}, \&
  {Armus}}]{2002ApJ...566..229C}
{Chary}, R., {Becklin}, E.~E., \& {Armus}, L. 2002, \apj, 566, 229

\bibitem[{{Christensen} {et~al.}(2004){Christensen}, {Hjorth}, \&
  {Gorosabel}}]{2004A&A...425..913C}
{Christensen}, L., {Hjorth}, J., \& {Gorosabel}, J. 2004, \aap, 425, 913

\bibitem[{{Coulter} {et~al.}(2017){Coulter}, {Foley}, {Kilpatrick}, {Drout},
  {Piro}, {Shappee}, {Siebert}, {Simon}, {Ulloa}, {Kasen}, {Madore},
  {Murguia-Berthier}, {Pan}, {Prochaska}, {Ramirez-Ruiz}, {Rest}, \&
  {Rojas-Bravo}}]{2017Sci...358.1556C}
{Coulter}, D.~A., {Foley}, R.~J., {Kilpatrick}, C.~D., {et~al.} 2017, Science,
  358, 1556

\bibitem[{{Currell} \& {Dowman}(2009)}]{odds}
{Currell}, G., \& {Dowman}, A. 2009, {Essential Mathematics and Statistics for
  Science}

\bibitem[{{Della Valle} {et~al.}(2006){Della Valle}, {Chincarini}, {Panagia},
  {Tagliaferri}, {Malesani}, {Testa}, {Fugazza}, {Campana}, {Covino},
  {Mangano}, {Antonelli}, {D'Avanzo}, {Hurley}, {Mirabel}, {Pellizza},
  {Piranomonte}, \& {Stella}}]{2006Natur.444.1050D}
{Della Valle}, M., {Chincarini}, G., {Panagia}, N., {et~al.} 2006, \nat, 444,
  1050

\bibitem[{{Eichler} {et~al.}(1989){Eichler}, {Livio}, {Piran}, \&
  {Schramm}}]{1989Natur.340..126E}
{Eichler}, D., {Livio}, M., {Piran}, T., \& {Schramm}, D.~N. 1989, \nat, 340,
  126

\bibitem[{{Faber} {et~al.}(2006){Faber}, {Baumgarte}, {Shapiro}, \&
  {Taniguchi}}]{2006ApJ...641L..93F}
{Faber}, J.~A., {Baumgarte}, T.~W., {Shapiro}, S.~L., \& {Taniguchi}, K. 2006,
  \apjl, 641, L93

\bibitem[{{Fong} \& {Berger}(2013)}]{2013ApJ...776...18F}
{Fong}, W., \& {Berger}, E. 2013, \apj, 776, 18

\bibitem[{{Fong} {et~al.}(2010){Fong}, {Berger}, \&
  {Fox}}]{2010ApJ...708....9F}
{Fong}, W., {Berger}, E., \& {Fox}, D.~B. 2010, \apj, 708, 9

\bibitem[{{Fox} {et~al.}(2005){Fox}, {Frail}, {Price}, {Kulkarni}, {Berger},
  {Piran}, {Soderberg}, {Cenko}, {Cameron}, {Gal-Yam}, {Kasliwal}, {Moon},
  {Harrison}, {Nakar}, {Schmidt}, {Penprase}, {Chevalier}, {Kumar}, {Roth},
  {Watson}, {Lee}, {Shectman}, {Phillips}, {Roth}, {McCarthy}, {Rauch},
  {Cowie}, {Peterson}, {Rich}, {Kawai}, {Aoki}, {Kosugi}, {Totani}, {Park},
  {MacFadyen}, \& {Hurley}}]{2005Natur.437..845F}
{Fox}, D.~B., {Frail}, D.~A., {Price}, P.~A., {et~al.} 2005, \nat, 437, 845

\bibitem[{{Fruchter} {et~al.}(2006){Fruchter}, {Levan}, {Strolger},
  {Vreeswijk}, {Thorsett}, {Bersier}, {Burud}, {Castro Cer{\'o}n},
  {Castro-Tirado}, {Conselice}, {Dahlen}, {Ferguson}, {Fynbo}, {Garnavich},
  {Gibbons}, {Gorosabel}, {Gull}, {Hjorth}, {Holland}, {Kouveliotou}, {Levay},
  {Livio}, {Metzger}, {Nugent}, {Petro}, {Pian}, {Rhoads}, {Riess}, {Sahu},
  {Smette}, {Tanvir}, {Wijers}, \& {Woosley}}]{2006Natur.441..463F}
{Fruchter}, A.~S., {Levan}, A.~J., {Strolger}, L., {et~al.} 2006, \nat, 441,
  463

\bibitem[{{Fryer} \& {Heger}(2005)}]{2005ApJ...623..302F}
{Fryer}, C.~L., \& {Heger}, A. 2005, \apj, 623, 302

\bibitem[{{Fynbo} {et~al.}(2006){Fynbo}, {Watson}, {Th{\"o}ne}, {Sollerman},
  {Bloom}, {Davis}, {Hjorth}, {Jakobsson}, {J{\o}rgensen}, {Graham},
  {Fruchter}, {Bersier}, {Kewley}, {Cassan}, {Castro Cer{\'o}n}, {Foley},
  {Gorosabel}, {Hinse}, {Horne}, {Jensen}, {Klose}, {Kocevski}, {Marquette},
  {Perley}, {Ramirez-Ruiz}, {Stritzinger}, {Vreeswijk}, {Wijers}, {Woller},
  {Xu}, \& {Zub}}]{2006Natur.444.1047F}
{Fynbo}, J.~P.~U., {Watson}, D., {Th{\"o}ne}, C.~C., {et~al.} 2006, \nat, 444,
  1047

\bibitem[{{Gal-Yam} {et~al.}(2006){Gal-Yam}, {Fox}, {Price}, {Ofek}, {Davis},
  {Leonard}, {Soderberg}, {Schmidt}, {Lewis}, {Peterson}, {Kulkarni}, {Berger},
  {Cenko}, {Sari}, {Sharon}, {Frail}, {Moon}, {Brown}, {Cucchiara}, {Harrison},
  {Piran}, {Persson}, {McCarthy}, {Penprase}, {Chevalier}, \&
  {MacFadyen}}]{2006Natur.444.1053G}
{Gal-Yam}, A., {Fox}, D.~B., {Price}, P.~A., {et~al.} 2006, \nat, 444, 1053

\bibitem[{{Galama} {et~al.}(1998){Galama}, {Vreeswijk}, {van Paradijs},
  {Kouveliotou}, {Augusteijn}, {B{\"o}hnhardt}, {Brewer}, {Doublier},
  {Gonzalez}, {Leibundgut}, {Lidman}, {Hainaut}, {Patat}, {Heise}, {in't Zand},
  {Hurley}, {Groot}, {Strom}, {Mazzali}, {Iwamoto}, {Nomoto}, {Umeda},
  {Nakamura}, {Young}, {Suzuki}, {Shigeyama}, {Koshut}, {Kippen}, {Robinson},
  {de Wildt}, {Wijers}, {Tanvir}, {Greiner}, {Pian}, {Palazzi}, {Frontera},
  {Masetti}, {Nicastro}, {Feroci}, {Costa}, {Piro}, {Peterson}, {Tinney},
  {Boyle}, {Cannon}, {Stathakis}, {Sadler}, {Begam}, \&
  {Ianna}}]{1998Natur.395..670G}
{Galama}, T.~J., {Vreeswijk}, P.~M., {van Paradijs}, J., {et~al.} 1998, \nat,
  395, 670

\bibitem[{{Gao} {et~al.}(2015){Gao}, {Ding}, {Wu}, {Dai}, \&
  {Zhang}}]{2015ApJ...807..163G}
{Gao}, H., {Ding}, X., {Wu}, X.-F., {Dai}, Z.-G., \& {Zhang}, B. 2015, \apj,
  807, 163

\bibitem[{{Gehrels} {et~al.}(2005){Gehrels}, {Sarazin}, {O'Brien}, {Zhang},
  {Barbier}, {Barthelmy}, {Blustin}, {Burrows}, {Cannizzo}, {Cummings}, {Goad},
  {Holland}, {Hurkett}, {Kennea}, {Levan}, {Markwardt}, {Mason}, {Meszaros},
  {Page}, {Palmer}, {Rol}, {Sakamoto}, {Willingale}, {Angelini}, {Beardmore},
  {Boyd}, {Breeveld}, {Campana}, {Chester}, {Chincarini}, {Cominsky},
  {Cusumano}, {de Pasquale}, {Fenimore}, {Giommi}, {Gronwall}, {Grupe}, {Hill},
  {Hinshaw}, {Hjorth}, {Hullinger}, {Hurley}, {Klose}, {Kobayashi},
  {Kouveliotou}, {Krimm}, {Mangano}, {Marshall}, {McGowan}, {Moretti},
  {Mushotzky}, {Nakazawa}, {Norris}, {Nousek}, {Osborne}, {Page}, {Parsons},
  {Patel}, {Perri}, {Poole}, {Romano}, {Roming}, {Rosen}, {Sato}, {Schady},
  {Smale}, {Sollerman}, {Starling}, {Still}, {Suzuki}, {Tagliaferri},
  {Takahashi}, {Tashiro}, {Tueller}, {Wells}, {White}, \&
  {Wijers}}]{2005Natur.437..851G}
{Gehrels}, N., {Sarazin}, C.~L., {O'Brien}, P.~T., {et~al.} 2005, \nat, 437,
  851

\bibitem[{{Gehrels} {et~al.}(2006){Gehrels}, {Norris}, {Barthelmy}, {Granot},
  {Kaneko}, {Kouveliotou}, {Markwardt}, {M{\'e}sz{\'a}ros}, {Nakar}, {Nousek},
  {O'Brien}, {Page}, {Palmer}, {Parsons}, {Roming}, {Sakamoto}, {Sarazin},
  {Schady}, {Stamatikos}, \& {Woosley}}]{2006Natur.444.1044G}
{Gehrels}, N., {Norris}, J.~P., {Barthelmy}, S.~D., {et~al.} 2006, \nat, 444,
  1044

\bibitem[{{Gendre} {et~al.}(2013){Gendre}, {Stratta}, {Atteia}, {Basa},
  {Bo{\"e}r}, {Coward}, {Cutini}, {D'Elia}, {Howell}, {Klotz}, \&
  {Piro}}]{2013ApJ...766...30G}
{Gendre}, B., {Stratta}, G., {Atteia}, J.~L., {et~al.} 2013, \apj, 766, 30

\bibitem[{{Goldstein} {et~al.}(2017){Goldstein}, {Veres}, {Burns}, {Briggs},
  {Hamburg}, {Kocevski}, {Wilson-Hodge}, {Preece}, {Poolakkil}, {Roberts},
  {Hui}, {Connaughton}, {Racusin}, {von Kienlin}, {Dal Canton}, {Christensen},
  {Littenberg}, {Siellez}, {Blackburn}, {Broida}, {Bissaldi}, {Cleveland},
  {Gibby}, {Giles}, {Kippen}, {McBreen}, {McEnery}, {Meegan}, {Paciesas}, \&
  {Stanbro}}]{2017ApJ...848L..14G}
{Goldstein}, A., {Veres}, P., {Burns}, E., {et~al.} 2017, \apjl, 848, L14

\bibitem[{{Granot} {et~al.}(2018){Granot}, {Gill}, {Guetta}, \& {De
  Colle}}]{2018MNRAS.481.1597G}
{Granot}, J., {Gill}, R., {Guetta}, D., \& {De Colle}, F. 2018, \mnras, 481,
  1597

\bibitem[{{Grupe} {et~al.}(2013){Grupe}, {Nousek}, {Veres}, {Zhang}, \&
  {Gehrels}}]{2013ApJS..209...20G}
{Grupe}, D., {Nousek}, J.~A., {Veres}, P., {Zhang}, B.-B., \& {Gehrels}, N.
  2013, \apjs, 209, 20

\bibitem[{{Hakkila} {et~al.}(2003){Hakkila}, {Giblin}, {Roiger}, {Haglin},
  {Paciesas}, \& {Meegan}}]{2003ApJ...582..320H}
{Hakkila}, J., {Giblin}, T.~W., {Roiger}, R.~J., {et~al.} 2003, \apj, 582, 320

\bibitem[{{Hand} \& {Yu}(2001)}]{hand2001}
{Hand}, D.~J., \& {Yu}, K. 2001, International Statistical Review, 69, 385

\bibitem[{{Hjorth} \& {Bloom}(2012)}]{2012grbu.book..169H}
{Hjorth}, J., \& {Bloom}, J.~S. 2012, {The Gamma-Ray Burst - Supernova
  Connection}, 169--190

\bibitem[{{Hjorth} {et~al.}(2003){Hjorth}, {Sollerman}, {M{\o}ller}, {Fynbo},
  {Woosley}, {Kouveliotou}, {Tanvir}, {Greiner}, {Andersen}, {Castro-Tirado},
  {Castro Cer{\'o}n}, {Fruchter}, {Gorosabel}, {Jakobsson}, {Kaper}, {Klose},
  {Masetti}, {Pedersen}, {Pedersen}, {Pian}, {Palazzi}, {Rhoads}, {Rol}, {van
  den Heuvel}, {Vreeswijk}, {Watson}, \& {Wijers}}]{2003Natur.423..847H}
{Hjorth}, J., {Sollerman}, J., {M{\o}ller}, P., {et~al.} 2003, \nat, 423, 847

\bibitem[{{Hjorth} {et~al.}(2005{\natexlab{a}}){Hjorth}, {Sollerman},
  {Gorosabel}, {Granot}, {Klose}, {Kouveliotou}, {Melinder}, {Ramirez-Ruiz},
  {Starling}, {Thomsen}, {Andersen}, {Fynbo}, {Jensen}, {Vreeswijk}, {Castro
  Cer{\'o}n}, {Jakobsson}, {Levan}, {Pedersen}, {Rhoads}, {Tanvir}, {Watson},
  \& {Wijers}}]{2005ApJ...630L.117H}
{Hjorth}, J., {Sollerman}, J., {Gorosabel}, J., {et~al.} 2005{\natexlab{a}},
  \apjl, 630, L117

\bibitem[{{Hjorth} {et~al.}(2005{\natexlab{b}}){Hjorth}, {Watson}, {Fynbo},
  {Price}, {Jensen}, {J{\o}rgensen}, {Kubas}, {Gorosabel}, {Jakobsson},
  {Sollerman}, {Pedersen}, \& {Kouveliotou}}]{2005Natur.437..859H}
{Hjorth}, J., {Watson}, D., {Fynbo}, J.~P.~U., {et~al.} 2005{\natexlab{b}},
  \nat, 437, 859

\bibitem[{{Horv{\'a}th}(1998)}]{1998ApJ...508..757H}
{Horv{\'a}th}, I. 1998, \apj, 508, 757

\bibitem[{{Horv{\'a}th} {et~al.}(2006){Horv{\'a}th}, {Bal{\'a}zs}, {Bagoly},
  {Ryde}, \& {M{\'e}sz{\'a}ros}}]{2006A&A...447...23H}
{Horv{\'a}th}, I., {Bal{\'a}zs}, L.~G., {Bagoly}, Z., {Ryde}, F., \&
  {M{\'e}sz{\'a}ros}, A. 2006, \aap, 447, 23

\bibitem[{{Horv{\'a}th} {et~al.}(2018){Horv{\'a}th}, {T{\'o}th}, {Hakkila},
  {T{\'o}th}, {Bal{\'a}zs}, {R{\'a}cz}, {Pint{\'e}r}, \&
  {Bagoly}}]{2018Ap&SS.363...53H}
{Horv{\'a}th}, I., {T{\'o}th}, B.~G., {Hakkila}, J., {et~al.} 2018, \apss, 363,
  53

\bibitem[{{Izzard} {et~al.}(2004){Izzard}, {Ramirez-Ruiz}, \&
  {Tout}}]{2004MNRAS.348.1215I}
{Izzard}, R.~G., {Ramirez-Ruiz}, E., \& {Tout}, C.~A. 2004, \mnras, 348, 1215

\bibitem[{{James} {et~al.}(2013){James}, {Witten}, {Hastie}, \&
  {Tibshirani}}]{James2013}
{James}, G., {Witten}, D., {Hastie}, T., \& {Tibshirani}, R. 2013, {An
  Introduction to Statistical Learning: with Applications in R}

\bibitem[{{Jin} {et~al.}(2016){Jin}, {Hotokezaka}, {Li}, {Tanaka}, {D'Avanzo},
  {Fan}, {Covino}, {Wei}, \& {Piran}}]{2016arXiv160307869J}
{Jin}, Z.-P., {Hotokezaka}, K., {Li}, X., {et~al.} 2016, ArXiv e-prints,
  arXiv:1603.07869

\bibitem[{{Kaneko} {et~al.}(2015){Kaneko}, {Bostanc{\i}}, {G{\"o}{\u g}{\"u}{\c
  s}}, \& {Lin}}]{2015MNRAS.452..824K}
{Kaneko}, Y., {Bostanc{\i}}, Z.~F., {G{\"o}{\u g}{\"u}{\c s}}, E., \& {Lin}, L.
  2015, \mnras, 452, 824

\bibitem[{{Kann} {et~al.}(2011){Kann}, {Klose}, {Zhang}, {Covino}, {Butler},
  {Malesani}, {Nakar}, {Wilson}, {Antonelli}, {Chincarini}, {Cobb}, {D'Avanzo},
  {D'Elia}, {Della Valle}, {Ferrero}, {Fugazza}, {Gorosabel}, {Israel},
  {Mannucci}, {Piranomonte}, {Schulze}, {Stella}, {Tagliaferri}, \&
  {Wiersema}}]{2011ApJ...734...96K}
{Kann}, D.~A., {Klose}, S., {Zhang}, B., {et~al.} 2011, \apj, 734, 96

\bibitem[{{Kennicutt} \& {Evans}(2012)}]{2012ARA&A..50..531K}
{Kennicutt}, R.~C., \& {Evans}, N.~J. 2012, \araa, 50, 531

\bibitem[{{Kewley} \& {Ellison}(2008)}]{2008ApJ...681.1183K}
{Kewley}, L.~J., \& {Ellison}, S.~L. 2008, \apj, 681, 1183

\bibitem[{{Kobulnicky} \& {Kewley}(2004)}]{2004ApJ...617..240K}
{Kobulnicky}, H.~A., \& {Kewley}, L.~J. 2004, \apj, 617, 240

\bibitem[{{Kouveliotou} {et~al.}(1993){Kouveliotou}, {Meegan}, {Fishman},
  {Bhat}, {Briggs}, {Koshut}, {Paciesas}, \& {Pendleton}}]{1993ApJ...413L.101K}
{Kouveliotou}, C., {Meegan}, C.~A., {Fishman}, G.~J., {et~al.} 1993, \apjl,
  413, L101

\bibitem[{{Kr{\"u}hler} {et~al.}(2015){Kr{\"u}hler}, {Malesani}, {Fynbo},
  {Hartoog}, {Hjorth}, {Jakobsson}, {Perley}, {Rossi}, {Schady}, {Schulze},
  {Tanvir}, {Vergani}, {Wiersema}, {Afonso}, {Bolmer}, {Cano}, {Covino},
  {D'Elia}, {de Ugarte Postigo}, {Filgas}, {Friis}, {Graham}, {Greiner},
  {Goldoni}, {Gomboc}, {Hammer}, {Japelj}, {Kann}, {Kaper}, {Klose}, {Levan},
  {Leloudas}, {Milvang-Jensen}, {Nicuesa Guelbenzu}, {Palazzi}, {Pian},
  {Piranomonte}, {S{\'a}nchez-Ram{\'{\i}}rez}, {Savaglio}, {Selsing},
  {Tagliaferri}, {Vreeswijk}, {Watson}, \& {Xu}}]{2015A&A...581A.125K}
{Kr{\"u}hler}, T., {Malesani}, D., {Fynbo}, J.~P.~U., {et~al.} 2015, \aap, 581,
  A125

\bibitem[{{Lazzati} {et~al.}(2001){Lazzati}, {Ramirez-Ruiz}, \&
  {Ghisellini}}]{2001A&A...379L..39L}
{Lazzati}, D., {Ramirez-Ruiz}, E., \& {Ghisellini}, G. 2001, \aap, 379, L39

\bibitem[{{Levan} {et~al.}(2014){Levan}, {Tanvir}, {Starling}, {Wiersema},
  {Page}, {Perley}, {Schulze}, {Wynn}, {Chornock}, {Hjorth}, {Cenko},
  {Fruchter}, {O'Brien}, {Brown}, {Tunnicliffe}, {Malesani}, {Jakobsson},
  {Watson}, {Berger}, {Bersier}, {Cobb}, {Covino}, {Cucchiara}, {de Ugarte
  Postigo}, {Fox}, {Gal-Yam}, {Goldoni}, {Gorosabel}, {Kaper}, {Kr{\"u}hler},
  {Karjalainen}, {Osborne}, {Pian}, {S{\'a}nchez-Ram{\'\i}rez}, {Schmidt},
  {Skillen}, {Tagliaferri}, {Th{\"o}ne}, {Vaduvescu}, {Wijers}, \&
  {Zauderer}}]{2014ApJ...781...13L}
{Levan}, A.~J., {Tanvir}, N.~R., {Starling}, R.~L.~C., {et~al.} 2014, \apj,
  781, 13

\bibitem[{{Levesque} {et~al.}(2010){Levesque}, {Bloom}, {Butler}, {Perley},
  {Cenko}, {Prochaska}, {Kewley}, {Bunker}, {Chen}, {Chornock}, {Filippenko},
  {Glazebrook}, {Lopez}, {Masiero}, {Modjaz}, {Morgan}, \&
  {Poznanski}}]{2010MNRAS.401..963L}
{Levesque}, E.~M., {Bloom}, J.~S., {Butler}, N.~R., {et~al.} 2010, \mnras, 401,
  963

\bibitem[{{Li} \& {Paczy{\'n}ski}(1998)}]{1998ApJ...507L..59L}
{Li}, L.-X., \& {Paczy{\'n}ski}, B. 1998, \apjl, 507, L59

\bibitem[{{Li} {et~al.}(2016){Li}, {Zhang}, \& {L{\"u}}}]{2016ApJS..227....7L}
{Li}, Y., {Zhang}, B., \& {L{\"u}}, H.-J. 2016, \apjs, 227, 7

\bibitem[{{Liang} {et~al.}(2007){Liang}, {Zhang}, {Virgili}, \&
  {Dai}}]{2007ApJ...662.1111L}
{Liang}, E., {Zhang}, B., {Virgili}, F., \& {Dai}, Z.~G. 2007, \apj, 662, 1111

\bibitem[{{L{\"u}} {et~al.}(2014){L{\"u}}, {Zhang}, {Liang}, {Zhang}, \&
  {Sakamoto}}]{2014MNRAS.442.1922L}
{L{\"u}}, H.-J., {Zhang}, B., {Liang}, E.-W., {Zhang}, B.-B., \& {Sakamoto}, T.
  2014, \mnras, 442, 1922

\bibitem[{{MacFadyen} \& {Woosley}(1999)}]{1999ApJ...524..262M}
{MacFadyen}, A.~I., \& {Woosley}, S.~E. 1999, \apj, 524, 262

\bibitem[{{Metzger} {et~al.}(2010){Metzger}, {Mart{\'{\i}}nez-Pinedo},
  {Darbha}, {Quataert}, {Arcones}, {Kasen}, {Thomas}, {Nugent}, {Panov}, \&
  {Zinner}}]{2010MNRAS.406.2650M}
{Metzger}, B.~D., {Mart{\'{\i}}nez-Pinedo}, G., {Darbha}, S., {et~al.} 2010,
  \mnras, 406, 2650

\bibitem[{{Mukherjee} {et~al.}(1998){Mukherjee}, {Feigelson}, {Jogesh Babu},
  {Murtagh}, {Fraley}, \& {Raftery}}]{1998ApJ...508..314M}
{Mukherjee}, S., {Feigelson}, E.~D., {Jogesh Babu}, G., {et~al.} 1998, \apj,
  508, 314

\bibitem[{{M{\"u}ller} \& {Guido}(2016)}]{machinelearning}
{M{\"u}ller}, C.~A., \& {Guido}, S. 2016, {Introduction to Machine Learning
  with Python: A Guide for Data Scientists}

\bibitem[{{Narayan} {et~al.}(1992){Narayan}, {Paczynski}, \&
  {Piran}}]{1992ApJ...395L..83N}
{Narayan}, R., {Paczynski}, B., \& {Piran}, T. 1992, \apjl, 395, L83

\bibitem[{{Norris} \& {Bonnell}(2006)}]{2006ApJ...643..266N}
{Norris}, J.~P., \& {Bonnell}, J.~T. 2006, \apj, 643, 266

\bibitem[{{Norris} {et~al.}(2010){Norris}, {Gehrels}, \&
  {Scargle}}]{2010ApJ...717..411N}
{Norris}, J.~P., {Gehrels}, N., \& {Scargle}, J.~D. 2010, \apj, 717, 411

\bibitem[{{Paczynski}(1986)}]{1986ApJ...308L..43P}
{Paczynski}, B. 1986, \apjl, 308, L43

\bibitem[{{Paczynski}(1991)}]{1991AcA....41..257P}
---. 1991, \actaa, 41, 257

\bibitem[{{Paczy{\'n}ski}(1998)}]{1998ApJ...494L..45P}
{Paczy{\'n}ski}, B. 1998, \apjl, 494, L45

\bibitem[{{Sahu} {et~al.}(1997){Sahu}, {Livio}, {Petro}, {Macchetto}, {van
  Paradijs}, {Kouveliotou}, {Fishman}, {Meegan}, {Groot}, \&
  {Galama}}]{1997Natur.387..476S}
{Sahu}, K.~C., {Livio}, M., {Petro}, L., {et~al.} 1997, \nat, 387, 476

\bibitem[{{Savaglio} {et~al.}(2009){Savaglio}, {Glazebrook}, \& {Le
  Borgne}}]{2009ApJ...691..182S}
{Savaglio}, S., {Glazebrook}, K., \& {Le Borgne}, D. 2009, \apj, 691, 182

\bibitem[{{Siemiginowska} {et~al.}(2019){Siemiginowska}, {Eadie}, {Czekala},
  {Feigelson}, {Ford}, {Kashyap}, {Kuhn}, {Loredo}, {Ntampaka}, {Stevens},
  {Avelino}, {Borne}, {Budavari}, {Burkhart}, {Cisewski-Kehe}, {Civano},
  {Chilingarian}, {van Dyk}, {Fabbiano}, {Finkbeiner}, {Foreman-Mackey},
  {Freeman}, {Fruscione}, {Goodman}, {Graham}, {Guenther}, {Hakkila},
  {Hernquist}, {Huppenkothen}, {James}, {Law}, {Lazio}, {Lee},
  {L{\'o}pez-Morales}, {Mahabal}, {Mandel}, {Meng}, {Moustakas}, {Muna},
  {Peek}, {Richards}, {Portillo}, {Scargle}, {de Souza}, {Speagle}, {Stassun},
  {Stenning}, {Taylor}, {Tremblay}, {Trimble}, {Yanamand ra-Fisher}, \&
  {Young}}]{2019BAAS...51c.355S}
{Siemiginowska}, A., {Eadie}, G., {Czekala}, I., {et~al.} 2019, \baas, 51, 355

\bibitem[{{Stanek} {et~al.}(2003){Stanek}, {Matheson}, {Garnavich}, {Martini},
  {Berlind}, {Caldwell}, {Challis}, {Brown}, {Schild}, {Krisciunas}, {Calkins},
  {Lee}, {Hathi}, {Jansen}, {Windhorst}, {Echevarria}, {Eisenstein}, {Pindor},
  {Olszewski}, {Harding}, {Holland}, \& {Bersier}}]{2003ApJ...591L..17S}
{Stanek}, K.~Z., {Matheson}, T., {Garnavich}, P.~M., {et~al.} 2003, \apjl, 591,
  L17

\bibitem[{{Tanvir} {et~al.}(2013){Tanvir}, {Levan}, {Fruchter}, {Hjorth},
  {Hounsell}, {Wiersema}, \& {Tunnicliffe}}]{2013Natur.500..547T}
{Tanvir}, N.~R., {Levan}, A.~J., {Fruchter}, A.~S., {et~al.} 2013, \nat, 500,
  547

\bibitem[{{Tsutsui} {et~al.}(2013){Tsutsui}, {Nakamura}, {Yonetoku},
  {Takahashi}, \& {Morihara}}]{2013PASJ...65....3T}
{Tsutsui}, R., {Nakamura}, T., {Yonetoku}, D., {Takahashi}, K., \& {Morihara},
  Y. 2013, \pasj, 65, 3

\bibitem[{{Villar} {et~al.}(2017){Villar}, {Guillochon}, {Berger}, {Metzger},
  {Cowperthwaite}, {Nicholl}, {Alexander}, {Blanchard}, {Chornock},
  {Eftekhari}, {Fong}, {Margutti}, \& {Williams}}]{2017ApJ...851L..21V}
{Villar}, V.~A., {Guillochon}, J., {Berger}, E., {et~al.} 2017, \apjl, 851, L21

\bibitem[{{Woosley}(1993)}]{1993ApJ...405..273W}
{Woosley}, S.~E. 1993, \apj, 405, 273

\bibitem[{{Woosley} \& {Bloom}(2006)}]{2006ARA&A..44..507W}
{Woosley}, S.~E., \& {Bloom}, J.~S. 2006, \araa, 44, 507

\bibitem[{{Xu} {et~al.}(2013){Xu}, {de Ugarte Postigo}, {Leloudas},
  {Kr{\"u}hler}, {Cano}, {Hjorth}, {Malesani}, {Fynbo}, {Th{\"o}ne},
  {S{\'a}nchez-Ram{\'{\i}}rez}, {Schulze}, {Jakobsson}, {Kaper}, {Sollerman},
  {Watson}, {Cabrera-Lavers}, {Cao}, {Covino}, {Flores}, {Geier}, {Gorosabel},
  {Hu}, {Milvang-Jensen}, {Sparre}, {Xin}, {Zhang}, {Zheng}, \&
  {Zou}}]{2013ApJ...776...98X}
{Xu}, D., {de Ugarte Postigo}, A., {Leloudas}, G., {et~al.} 2013, \apj, 776, 98

\bibitem[{{Yang} {et~al.}(2015){Yang}, {Jin}, {Li}, {Covino}, {Zheng},
  {Hotokezaka}, {Fan}, {Piran}, \& {Wei}}]{2015NatCo...6E7323Y}
{Yang}, B., {Jin}, Z.-P., {Li}, X., {et~al.} 2015, Nature Communications, 6,
  7323

\bibitem[{{Zhang}(2006)}]{2006Natur.444.1010Z}
{Zhang}, B. 2006, \nat, 444, 1010

\bibitem[{{Zhang}(2018)}]{Zhang2018book}
---. 2018, {The Physics of Gamma-Ray Bursts}, doi:10.1017/9781139226530

\bibitem[{{Zhang} {et~al.}(2007){Zhang}, {Zhang}, {Liang}, {Gehrels},
  {Burrows}, \& {M{\'e}sz{\'a}ros}}]{2007ApJ...655L..25Z}
{Zhang}, B., {Zhang}, B.-B., {Liang}, E.-W., {et~al.} 2007, \apjl, 655, L25

\bibitem[{{Zhang} {et~al.}(2009){Zhang}, {Zhang}, {Virgili}, {Liang}, {Kann},
  {Wu}, {Proga}, {Lv}, {Toma}, {M{\'e}sz{\'a}ros}, {Burrows}, {Roming}, \&
  {Gehrels}}]{2009ApJ...703.1696Z}
{Zhang}, B., {Zhang}, B.-B., {Virgili}, F.~J., {et~al.} 2009, \apj, 703, 1696

\bibitem[{{Zhang} {et~al.}(2014){Zhang}, {Zhang}, {Murase}, {Connaughton}, \&
  {Briggs}}]{2014ApJ...787...66Z}
{Zhang}, B.-B., {Zhang}, B., {Murase}, K., {Connaughton}, V., \& {Briggs},
  M.~S. 2014, \apj, 787, 66

\bibitem[{{Zhang} {et~al.}(2018){Zhang}, {Zhang}, {Sun}, {Lei}, {Gao}, {Li},
  {Shao}, {Zhao}, {Hu}, {L{\"u}}, {Wu}, {Fan}, {Wang}, {Castro-Tirado},
  {Zhang}, {Yu}, {Cao}, \& {Liang}}]{2018NatCo...9..447Z}
{Zhang}, B.-B., {Zhang}, B., {Sun}, H., {et~al.} 2018, Nature Communications,
  9, 447

\end{thebibliography}

\pagebreak

\begin{appendix}

\section{Analytical Results of Maximum Likelihood Estimations}

\subsection{Gaussian Distribution}

For a Gaussian distribution 
$$f(x|\mu, \sigma^2)=\frac{1}{\sqrt{2\pi \sigma^2}}{\rm exp}\big( - \frac{(x-\mu)^2}{2\sigma^2}\big),$$
the logarithmic likelihood for a sample with $n$ variables is 
$${\rm ln}\ \mathcal{L}= {\rm ln} \prod_{i=1}^n f(x_i|\mu, \sigma^2)= {\rm ln} \prod_{i=1}^n \frac{1}{\sqrt{2\pi \sigma^2}}{\rm exp}\big( - \frac{(x_i-\mu)^2}{2\sigma^2}\big)= -\frac{n}{2} {\rm ln} (2\pi \sigma^2)-\frac{1}{2\sigma^2} \sum_{i=1}^n (x_i-\mu)^2. $$

To obtain the extreme of the logarithmic likelihood, thus the extreme of the likelihood, we require the derivatives of this log-likelihood to be zero
$$\frac{\partial}{\partial \mu}({\rm ln} \mathcal{L})
=-\frac{2(\sum_{i=1}^n x_i-n \mu)}{2\sigma^2}=0 
\Rightarrow \mu=\frac{\sum_{i=1}^n x_i}{n}=\bar{x}$$

$$\frac{\partial}{\partial \sigma} ({\rm ln} \mathcal{L}) =
-\frac{n}{\sigma}+\frac{\sum_{i=1}^n(x_i-\mu)^2}{\sigma^3}
\Rightarrow \sigma^2=\frac{\sum_{i=1}^n (x_i-\mu)^2}{n}.$$

\subsection{Exponential Distribution}

The exponential distribution we used for the parameter light fraction $F_{\rm light}$ is $f(x|\gamma)=A {\rm exp}(\gamma x)$. To have the integrated value within $[0,1]$ normalized to be $1$, one has
$$f(x|\gamma)=A {\rm exp}(\gamma x), A=\frac{\gamma}{{\rm exp}(\gamma)-1}.$$
The logarithmic likelihood for a sample with $n$ variables is 
$${\rm ln}\ \mathcal{L}= {\rm ln} \prod_{i=1}^n f(x_i|\gamma)= 
{\rm ln} \prod_{i=1}^n A {\rm exp} (\gamma x_i)=n{\rm ln}A+\gamma \sum_{i=1}^n x_i. $$

To obtain the extreme of the logarithmic likelihood, thus the extreme of the likelihood, we require the derivatives of this log-likelihood to be zero
$$\frac{\partial}{\partial \gamma}({\rm ln} \mathcal{L})
=n\frac{e^{\gamma}-1-\gamma e^{\gamma}}{\gamma(e^{\gamma}-1)}+\sum_{i=1}^n x_i=0
\Rightarrow \bar{x}=\frac{1}{1-e^{-\gamma}}-\frac{1}{\gamma}$$

\end{appendix}

\end{document}